\documentclass[journal]{IEEEtran}

\usepackage{booktabs}
\usepackage{threeparttable}
\usepackage{amsmath,graphicx,cite,amssymb}
\usepackage{amsfonts,multirow,bm,array,setspace,stfloats}
\usepackage{xcolor}
\usepackage{xpatch}

\usepackage{graphicx,float,cite,amssymb}
\usepackage{graphics}
 \DeclareGraphicsExtensions{.pdf,.jpeg,.png,.jpg}   
\usepackage{multirow,bm,bbm,array,setspace}
\usepackage{textcomp}
\usepackage{psfrag}
\usepackage{color}
\usepackage{pstricks,enumerate}
\usepackage{bbm}
\usepackage{subfigure}

\usepackage{algorithm,algpseudocode}
\makeatletter

\def\changeBibColor#1{%
  \in@{#1}{grouping_Zhangrui_LWC_2020,ZhangRui_ICC_2020,grouping_Fang_TWC_2020,grouping_Zhangrui_TCOM_2020,beamtraining_Zhangw_2021,beamtraining_LiH_2022,beamtraining_FangJ_TVT,beamtraing_Zhangrui_2020,Nadeem_WCOM,Psomas_WCOM,TwoTime}
  \ifin@\color{black}\else\normalcolor\fi
}
 
\xpatchcmd\@bibitem
  {\item}
  {\changeBibColor{#1}\item}
  {}{\fail}
 
\xpatchcmd\@lbibitem
  {\item}
  {\changeBibColor{#2}\item}
  {}{\fail}
\makeatother

\newcommand{\distas}[1]{\mathbin{\overset{#1}{\kern\z@\sim}}}%
\newsavebox{\mybox}\newsavebox{\mysim}
\newcommand{\distras}[1]{%
  \savebox{\mybox}{\hbox{\kern1pt$\scriptstyle#1$\kern1pt}}%
  \savebox{\mysim}{\hbox{$\sim$}}%
  \mathbin{\overset{#1}{\kern\z@\resizebox{\wd\mybox}{\ht\mysim}{$\sim$}}}%
}
\makeatother
\ifCLASSINFOpdf
\else
\fi

\newtheorem{proposition}{Proposition}

\newtheorem{lemma}{Lemma}
\newtheorem{theorem}{Theorem}
\newtheorem{corollary}{Corollary}

\newtheorem{remark}{Remark}

\newcommand{\bA}{\bm A}
\newcommand{\ba}{\bm a}

\newcommand{\bx}{\bm{x}}

\newcommand{\bs}{\bm{s}}
\newcommand{\bI}{\bm{I}}

\newcommand{\bv}{\bm v}

\newcommand{\bu}{\bm u}

\newcommand{\bW}{\bm{W}}

\newcommand{\btheta}{\bm \theta}

\begin{document}
%
\title{Configuring Intelligent Reflecting Surface with Performance Guarantees: Blind Beamforming}
\author{
	\IEEEauthorblockN{
	Shuyi Ren, \IEEEmembership{Student Member,~IEEE}, Kaiming Shen, \IEEEmembership{Member,~IEEE},\\ 
	Yaowen Zhang, \IEEEmembership{Student Member,~IEEE}, Xin Li, Xin Chen, and Zhi-Quan Luo, \IEEEmembership{Fellow,~IEEE}
} 
\thanks{Manuscript submitted to IEEE Transactions on Wireless Communications 4 December, 2021; accepted 20 October 2022. The work of S. Ren, K. Shen, and Y. Zhang was supported in part by the National Natural Science Foundation of China (NSFC) under Grant 92167202, in part by the NSFC under Grant 62001411, and in part by the Huawei Technologies. The work of Z.-Q. Luo was supported in part by the NSFC under Grant 61731018 and in part by the Guangdong Provincial Key Laboratory of Big Data Computing. Part of this article was presented at the IEEE Global Communications Conference (GLOBECOM), Madrid, Spain, 2021 [DOI: 10.1109/GLOBECOM46510.2021.9685790] \emph{(Corresponding author: Kaiming Shen.)}

S. Ren, K. Shen, and Y. Zhang are with the School of Science and Engineering, The Chinese University of Hong Kong, Shenzhen, 518172, China (e-mail: shuyiren@link.cuhk.edu.cn; shenkaiming@cuhk.edu.cn; yaowenzhang@link.cuhk.edu.cn). 

X. Li and X. Chen are with Huawei Technologies (e-mail: razor.lixin@huawei.com; chenxin@huawei.com).

Z.-Q. Luo is both with the School of Science and Engineering, The Chinese University of Hong Kong, Shenzhen, 518172, China and with Shenzhen Research Institute of Big Data, China (e-mail: luozq@cuhk.edu.cn).}
}

%


\maketitle

\begin{abstract}
This paper proposes a blind beamforming strategy for intelligent reflecting surface (IRS), aiming to boost the signal-to-noise ratio (SNR) by coordinating phase shifts across the reflective elements in the absence of channel information. Differing from most existing approaches that first estimate channels and then optimize phase shifts, the proposed blind beamforming method explores the wireless environment by extracting statistical features directly from random samples of the received signal power, without acquiring channel station information (CSI). This new method just requires a polynomial number of random samples to provide a quadratic SNR boost in the number of reflective elements without CSI, whereas the standard random-max sampling algorithm can only achieve a linear boost under the same condition. Moreover, we interpret blind beamforming from a least-squares point of view. Field tests demonstrate the significant advantages of the proposed blind beamforming approach over the benchmark methods in enhancing wireless transmission.
\end{abstract}


\section{Introduction}
\label{sec:overview}

\IEEEPARstart{T}{he main problem} in configuring intelligent reflecting surface (IRS) is that of coordinating phase shifts across the reflective elements, namely passive beamforming, in order to boost the data throughput and transmission reliability in a volatile wireless environment (e.g., in an automated industrial factory). While the traditional paradigm of first estimating channels and then optimizing phase shifts has been extensively considered for the IRS beamforming in the literature to date, there is a surge of research interests in \emph{blind beamforming} without any channel knowledge. This work aims at a statistical approach to blind beamforming, which guarantees a quadratic boost of signal-to-noise ratio (SNR) by using only a polynomial number of random samples of the received signal power.
Aside from theoretical justifications, field tests in a real-world environment validate the practical effectiveness of the proposed blind beamforming method.



By treating IRS as an unsourced multi-antenna device, the existing works mostly adopt the conventional model-driven beamforming approach based on channel estimation. However, 
channel acquisition for IRS-aided systems poses formidable challenges in engineering practice, mainly recognized in the following three respects: 
\begin{enumerate}[i.]
	\item Each reflected channel alone can be easily overwhelmed by the much stronger background channel and noise, so it is difficult to recover the reflected channels precisely.
	\item We must modify the current networking protocol to enable channel estimation for IRS, e.g., it entails acquiring the in-phase and quadrature components from communication chips but this is not supported by the current 5G protocol.
	\item Channel estimation would increase the computational complexity of the IRS configuration, e.g., it typically involves the matrix inversion.
\end{enumerate}
Actually, the authors of \cite{Balakrishnan_NSDI20} also realized the above issues when developing their IRS prototype. Like \cite{Balakrishnan_NSDI20}, this work suggests removing channel estimation altogether from the IRS configuration. In particular, one must distinguish between the deep learning approach in \cite{Zhang_unsupervised_AI_20,Yuan_AI_21,yuen_reinforcement_AI_21,Wen_reinforcement_AI_20,wy_AI_21} and our statistical approach with performance guarantees, albeit both are data-driven.

There has been an explosion of research activities in the IRS field over the past few years. As summarized in \cite{Liang_survey_20}, semidefinite relaxation (SDR) \cite{luo_SDR} and fractional programming (FP) \cite{FP1,FP2} constitute two popular optimization tools in this area, e.g., \cite{rui_SDP_BF_19,Johansson_est_19,rui_double_BF_21,Nallanathan_SDP_BF_20,Dobre_SDP_BF_21,YF_maxmin_BF_21,Skoglund_decentralized_BF_21} find SDR useful because of the quadratic form of the IRS beamforming problem, while \cite{Ni_FP_BF_20,Chen_FP_BF_21,Ding_FP_BF_21,Hossain_FP_BF_21,Ni_Lin_FP_BF_21,poor_active_BF_21,Dai_FP_BF_21,Pei_FP_BF_21} rely on the FP technique to deal with the signal-to-noise-plus-interference ratio (SINR) terms. Other optimization approaches in the existing literature include successive convex approximation \cite{Dhahir_SCA_BF_20,mm_2timescale_BF_21}, minorization-maximization algorithm \cite{Ng_impairment_BF_21}, and alternating direction method of multipliers (ADMM) \cite{fang_SVD_BF_20,Skoglund_decentralized_BF_21}.

Nonetheless, a real challenge facing all the above model-driven optimization methods is that their performance heavily depends upon the accurate information of channels---which is quite costly to acquire in practice because of the three reasons as noted earlier. A simple idea from \cite{Johansson_est_19} is to sequentially activate each individual reflective element at a time to measure the corresponding reflected channel. Apart from being inefficient, such channel estimation is prone to large error since each reflected channel alone can be much weaker than the background channel. The more recent works \cite{grouping_Zhangrui_TCOM_2020,grouping_Fang_TWC_2020} promote this {\footnotesize{ON-OFF}} policy by activating a group of reflective elements simultaneously. 
Another line of studies \cite{grouping_Zhangrui_LWC_2020,Carvalho_est_20,Boyer_est_21,Alouini_est_20} focus on the pilot sequence design by means of the discrete Fourier transform (DFT) matrix; the Cram\'{e}r-Rao lower bound to the least squares channel estimation can be attained under certain condition \cite{Carvalho_est_20}. Moreover, \cite{Li_CS_est_20,Alouini_Renzo_AI_20,Debbah_Zhang_est_21,Yu_est_20} view the channel estimation for IRS as a compressed sensing task. All the above methods only have been tested in simulations with artificially generated channels, but the real-world wireless environment can be far more chaotic according to our field tests.

The idea of configuring IRS without channel estimation has worked its way into a new frontier.
Both \cite{Psomas_WCOM} and \cite{Nadeem_WCOM} advocate a random rotation strategy for passive beamforming, thereby not requiring instantaneous channel information. Moreover, \cite{beamtraing_Zhangrui_2020,beamtraining_FangJ_TVT,beamtraining_LiH_2022,beamtraining_Zhangw_2021} suggest simply sweeping all possible directions of the reflected beam; as a consequence not any channel information is required. But this beam training approach is limited to the millimeter/terahertz waves with sharp beams. Deep learning is another approach. As opposed to \cite{Alouini_Renzo_AI_20,Chatzinotas_AI_20,Zhang_unsupervised_AI_20,Yuan_AI_21,yuen_reinforcement_AI_21,Wen_reinforcement_AI_20} that apply neural networks to either channel estimation or passive beamforming for IRS, the recent work \cite{wy_AI_21} proposes combining the two stages into directly learning how to perform beamforming based on the received pilot signal. It is argued in \cite{wy_AI_21} that such joint deep learning is capable of extracting more pertinent information from the raw data. Differing from these neural net-based approaches, the method called RFocus in \cite{Balakrishnan_NSDI20} uses the statistics of the received signal. Inspired by RFocus, this work also pursues a statistical approach and shows that the proposed blind beamforming method can strike even better provable performance.

The primary idea behind the proposed blind beamforming method is to exploit the conditional sample mean of the received signal power. We further analyze how the random sample size affects the performance of blind beamforming. The main contributions of this work are summarized as follows:

\begin{itemize}
\item We take a fundamental look at the conventional baseline method of random-max sampling, showing that its SNR boost is quasilinear in the number of reflective elements.
\item We propose a novel idea of using the conditional sample mean of the received signal power to optimize phase shifts and thus avoid channel estimation. When there are more than two phase shift choices for each reflective element, the proposed conditional sample mean algorithm yields a quadratic SNR boost in the number of reflective elements---which is the highest possible SNR boost.
\item We interpret the conditional sample mean method from a least-squares perspective; this interpretation further leads us to a generic family of the blind beamforming approaches.
\item The proposed blind beamforming method is also modified to account for multiple users and multiple antennas, by introducing the conditional sample mean of the general utility function.
\item We further enhance the proposed blind beamforming based on a constant-factor approximation algorithm in \cite{IRS:opt_21}. The resulting enhanced conditional sample mean algorithm yields better robustness.

\end{itemize}



The rest of the paper is organized as follows. Section \ref{sec:model} describes the system model and problem formulation. Section \ref{sec:withoutCSI} examines the different approaches to blind beamforming: We start with two common heuristic methods, followed by the proposed blind beamforming algorithm with provable better performance; we also provide a least-squares interpretation of this new method and devise its extensions to multiple antennas and multiple users. Section \ref{sec:ECSM} further uses a constant-factor approximation technique to improve the robustness of the proposed method. {Section \ref{sec:experiments}} presents the field tests and simulation results. Finally, Section \ref{sec:conclusion} concludes the paper.
 
Throughout the paper, we use the bold lower-case letter to denote a vector, the bold upper-case letter a matrix, and the calligraphy upper-case letter a set. For a matrix $\bA$, $\bA^\mathsf{T}$ refers to the transpose, $\bA^\mathsf{H}$ the conjugate transpose, and $\bA^{-1}$ the inverse. For a vector $\ba$, $\|\ba\|$ refers to the Euclidean norm, $\ba^\mathsf{T}$ the transpose, and $\ba^\mathsf{H}$ the conjugate transpose. The cardinality of a set $\mathcal Q$ is denoted as $|\mathcal Q|$. The set of real numbers is denoted as $\mathbb R$. The set of complex numbers is denoted as $\mathbb C$. For a complex number $u$, $\mathrm{Re}\{u\}$, $\mathrm{Im}\{u\}$, and $\mathrm{Arg}(u)$ refer to the real part, the imaginary part, and the principal argument of $u$, respectively. For an event $\mathcal E$, let $\mathbb{P}\{\mathcal E\}$ be its probability and let $\mathcal E^c$ be its complement. Let $\mathbb E[X]$ be the expectation of the random variable $X$, and let $\widehat{\mathbb E}[X]$ be the sample mean. We use the Bachmann-Landau notation: $f(n)=O(g(n))$ if there exists some $c>0$ such that $|f(n)|\le cg(n)$ for $n$ sufficiently large; $f(n)=o(g(n))$ if there exists some $c>0$ such that $|f(n)|< cg(n)$ for $n$ sufficiently large; $f(n)=\Omega(g(n))$ if there exists some $c>0$ such that $f(n)\ge cg(n)$ for $n$ sufficiently large; $f(n)=\Theta(g(n))$ if $f(n)=O(g(n))$ and $f(n)=\Omega(g(n))$ both hold.

\section{System Model}
\label{sec:model}

Consider a pair of transmitter and receiver, along with an IRS deployed to facilitate the data transmission between them. (The multiple-user case will be discussed in Section \ref{subsec:MIMO}.) The IRS consists of $N$ reflective elements. Let $h_n\in\mathbb C$, $n=1,\ldots,N$, be the cascaded channel from the transmitter to the receiver that is induced by the $n$th reflective element; let $h_0\in\mathbb C$ be the superposition of the rest channels independent of the IRS, namely the \emph{background channel}. Each channel can be rewritten in a polar form as
\begin{equation}
\label{hn}
	h_n= \beta_ne^{j\alpha_n},\;\;n=0,\ldots,N,
\end{equation} 
with the magnitude $\beta_n\in(0,1)$ and the phase $\alpha_n\in[0,2\pi)$.

Denote the IRS beamformer by $\bm\theta=(\theta_1,\ldots,\theta_N)$, with each $\theta_n$ representing the phase shift of the $n$th reflective element. The choice of each $\theta_n$ is restricted to the discrete set
\begin{equation}
\label{Phi}
	\Phi_K= \big\{\omega,2\omega,\ldots,K\omega\big\},
\end{equation}
wherein the distance $\omega$ is given by
\begin{equation}
\label{omega}
	\omega=\frac{2\pi}K.
\end{equation}
Let $X\in\mathbb C$ be the transmit signal with the mean power $P$, i.e., $\mathbb E[|X|^2]=P$. The received signal $Y\in\mathbb C$ can be computed as
\begin{align}
    Y &= \Bigg(h_0+\sum^N_{n=1}h_ne^{j\theta_n}\Bigg)X+Z,
\end{align}
where an i.i.d. random variable $Z\sim\mathcal{CN}(0,\sigma^2)$ models the additive background noise. The received SNR amounts to
\begin{subequations}
\begin{align}
\label{snr}
	\mathsf{SNR} &=\frac{\mathbb E[|Y-Z|^2]}{\mathbb E[|Z|^2]}\\
	&=\frac{P}{\sigma^2}\left|\beta_0e^{j\alpha_0}+\sum^N_{n=1}\beta_ne^{j(\alpha_n+\theta_n)}\right|^2.
\end{align}
\end{subequations}
In particular, the baseline SNR without reflection channels is
\begin{equation}
	\mathsf{SNR}_0 = \frac{P\beta^2_0}{\sigma^2}.
\end{equation}
The \emph{SNR boost} is then defined as
\begin{subequations}
\begin{align}
\label{SNR_boost}
	f(\bm\theta) &= \frac{\mathsf{SNR}}{\mathsf{SNR}_0}\\
	&=\frac{1}{\beta^2_0}\left|\beta_0e^{j\alpha_0}+\sum^N_{n=1}\beta_ne^{j(\alpha_n+\theta_n)}\right|^2.
\end{align}
\end{subequations}
We seek the optimal $\bm\theta$ that maximizes the SNR boost, i.e.,
\begin{subequations}
\label{prob:snr}
\begin{align}
&\underset{\bm\theta}{\text{maximize}} \quad f(\bm\theta)\\
&\text{subject to}\quad \theta_n\in\Phi_K\;\text{for all}\; n=1,\ldots,N.
\end{align}
\end{subequations}    

The challenges in solving the above problem are two-fold. First, it is numerically difficult to optimize the discrete variable $\btheta$. Second, from a practical standpoint, it is costly to obtain the channel information $\{h_0,\ldots,h_N\}$. Notice that this work does not assume coordination between base-station and IRS, while some existing works \cite{rui_SDP_BF_19,poor_active_BF_21,Dai_FP_BF_21,Dhahir_SCA_BF_20,mm_2timescale_BF_21} consider optimizing base-station and IRS jointly.



\section{Blind Beamforming}
\label{sec:withoutCSI}


We start by defining the \emph{average per element reflection power gain} as
\begin{equation}
	\bar\beta^2=\frac{1}{N}\sum^N_{n=1}\beta^2_n.
\end{equation}
An upper order bound on the achievable SNR boost can be readily obtained as stated in the following proposition.
\begin{proposition}[Converse]
\label{prop:upper_bound}
	The SNR boost is at most quadratic in the number of reflective elements, i.e.,
	\begin{equation}
		f(\btheta)=\frac{\bar\beta^2}{\beta^2_0}\cdot O(N^2).
	\end{equation}
	We remark that a similar result has been shown in the earlier work \cite{rui_SDP_BF_19}.
\end{proposition}
\begin{IEEEproof}
	The main idea follows \cite{rui_SDP_BF_19} closely. In the ideal case, every $h_ne^{j\theta_n}$ can be aligned with $h_0$ exactly, so the SNR boost is upper bounded as $f(\btheta)=1/\beta^2_0\cdot(\beta_0+\sum^N_{n=1}\beta_n)^2\le (N+1)/\beta^2_0\cdot(\beta^2_0+N\bar\beta^2)=\bar\beta^2/\beta^2_0\cdot O(N^2)$.
\end{IEEEproof}

While \cite{rui_SDP_BF_19} shows that the above upper bound can be achieved given channel state information (CSI), we show that it is achievable even without CSI. Before proceeding to this main result, we first look at two common heuristic methods in the next subsection.


\subsection{Two Common Heuristics}
\label{subsec:heuristics}

\subsubsection{Closest Point Projection (CPP)}

As the number of phase shift choices $K\rightarrow\infty$, the discrete beamforming problem in (\ref{prob:snr}) reduces to the continuous. Consequently, the ideal case assumed in the proof of Proposition \ref{prop:upper_bound} can now be realized, i.e., we get all the $h_ne^{j\theta_n}$ lined up exactly in the same direction as $h_0$. Specifically, with the phase difference between $h_0$ and $h_n$ denoted by
\begin{equation}
\label{Delta}
	\Delta_n = \alpha_0-\alpha_n,
\end{equation}
each $\theta_n$ is optimally set to $\Delta_n$.
To tackle the discrete case with a finite $K$, a natural idea \cite{Gao_DBF_21,Qian_DBF_21,you_zheng_zhang_jsac20,Schober_DBF_21,Smith_DBF_21,Yuen_DBF_20,wu_zhang_TCOM20} is to round the continuous solution to the closest point in the discrete set $\Phi_K$, i.e.,
\begin{equation}
\label{greedy_alg}
	{\theta}^{\text{CPP}}_n = \arg\min_{\theta_n\in\Phi_K}\big|\theta_n-\Delta_n\big|.
\end{equation}
Notice that channel information is required in the above procedure.

\subsubsection{Random-Max Sampling Algorithm (RMS)}

In order to get rid of channel information, we could just try out $T$ random samples of $\btheta$ and then choose the best. Let $t=1,\ldots,T$ be the index of random sample; let $\theta_{nt}$ be the phase shift of the $n$th reflective element in the $t$th random sample and let $\btheta_t=(\theta_{1t},\ldots,\theta_{Nt})$. Each $\theta_{nt}$ is drawn from the set $\Phi_K$ uniformly and independently. The received signal of the $t$th random sample is given by
\begin{equation}
\label{yt}
	Y_t = \left(h_0+\sum^N_{n=1}h_ne^{j\theta_{nt}}\right) X_t+Z_t,
\end{equation}
where the training signal $X_t$ has a constant power $P$ across $t$. The RMS method decides the entire vector $\btheta$ according to the received signal power, i.e.,
\begin{equation}
\label{RMS}
	\btheta^{\text{RMS}} = \btheta_{t^\star}\;\;\text{where}\;\;t^\star=\arg\max_{1\le t\le T} |Y_t|^2.
\end{equation}
It can be seen that the performance of RMS is closely related to the number of random samples $T$. In the following theorem, we examine this relationship by establishing the scaling law of SNR boost for RMS.


\begin{theorem}
\label{theorem:random}
Consider $T$ i.i.d. random samples of $\btheta$ uniformly drawn over $\Phi_K$. The expected SNR boost achieved by the RMS method in \eqref{RMS} has the following order bounds:
\begin{align}
	\mathbb E\big[f(\btheta^\text{RMS})\big] &= \frac{\bar\beta^2}{\beta^2_0}\cdot\Omega(N\log T)\;\;\text{if}\;\; T = o(\sqrt{N}),
	\label{theorem:random:lower}\\
	\mathbb E\big[f(\btheta^\text{RMS})\big] &= \frac{\bar\beta^2}{\beta^2_0}\cdot O(N\log T)\;\;\text{in general},
	\label{theorem:random:upper}
\end{align}
where the expectation is taken over random samples of $\btheta$.
\end{theorem}
\begin{IEEEproof}
	See Appendix \ref{proof:RMS}.
\end{IEEEproof}

Notice that (\ref{theorem:random:lower}) is a lower bound while (\ref{theorem:random:upper}) is an upper bound, so combining them immediately leads us to a tight order bound, as stated in the corollary below.
\begin{corollary}
	If $T=o(\sqrt{N})$, the expected SNR boost achieved by RMS meets $\mathbb E\big[f(\btheta^\text{RMS})\big]=(\bar\beta^2/\beta^2_0)\cdot\Theta(N\log T)$.
\end{corollary}

Because $T$ is polynomial in $N$ in most practical cases, the scaling rate $\Theta(N\log T)\approx\Theta(N)$, i.e., RMS yields a quasilinear SNR boost. Hence, there is gap between this achievability and the converse in Proposition \ref{prop:upper_bound}. The main drawback of RMS is that it solely relies on the ``best'' random sample whilst neglecting the rest. To close the gap between the achievability and converse of the SNR boost, we propose a statistical approach in the next subsection.


\subsection{Proposed Conditional Sample Mean (CSM) Method}

We still generate a total of $T$ random samples of $\btheta$ in an i.i.d. fashion\footnote{To improve the efficiency of sampling in practice, we should draw $\bm\theta$ from a codebook rather than randomly generate it on the fly.}. Use $\mathcal Q_{nk}\subseteq\{1,\ldots,T\}$ to denote the subset of all those random samples with $\theta_{nt}=k\omega$, i.e.,
\begin{equation}
\label{Q_nk}
	\mathcal Q_{nk} = \{t:\theta_{nt}=k\omega\}.
\end{equation}
We then compute the sample mean of the received signal power $|Y_t|^2$ conditioned on each $\mathcal Q_{nk}$:
\begin{equation}
	\label{conditional_mean}
	\widehat{\mathbb E}[|Y|^2|\theta_n=k\omega] = \frac{1}{|\mathcal Q_{nk}|}\sum_{t\in\mathcal Q_{nk}}|Y_t|^2.
\end{equation}
Intuitively, the conditional sample mean $\widehat{\mathbb E}[|Y|^2|\theta_n=k\omega]$ characterizes the average performance of setting $\theta_n=k\omega$. Thus, it is a natural idea to choose each $\theta_n$ according to the conditional sample mean, i.e.,
\begin{equation}
	\label{blind:CSM}
	{\theta}^{\text{CSM}}_n=\arg\max_{\varphi\in\Phi_K} \widehat{\mathbb E}[|Y|^2|\theta_n=\varphi].
\end{equation}
The above \emph{Conditional Sample Mean (CSM)} method is summarized in Algorithm \ref{alg:BLD}. The following theorem analyzes the training cost of CSM.

\begin{algorithm}[t]
\caption{Conditional Sample Mean (CSM)}
\label{alg:BLD}
\begin{algorithmic}[1]
\State\textbf{input:} $\Phi_K$, $N$, $T$;
\For{$t=1,2,\ldots,T$}
	\State generate $\btheta_{t}=(\theta_{1t},\ldots,\theta_{Nt})$ i.i.d. based on $\Phi_k$;
	\State measure received signal power $|Y_t|^2$ with $\btheta_{t}$;
\EndFor
\For{$n=1,2,\ldots,N$}
	\For{$k=1,2,\ldots,K$}
		\State compute $\widehat{\mathbb E}[|Y|^2|\theta_n=k\omega]$ according to (\ref{conditional_mean});
	\EndFor
	\State decide $\theta^\text{CSM}_n$ according to (\ref{blind:CSM});
\EndFor
\State\textbf{output:} $\btheta^\text{CSM}=(\theta^\text{CSM}_1,\ldots,\theta^\text{CSM}_N)$.
\end{algorithmic}
\end{algorithm}

\begin{theorem}
	\label{theorem:blind}
	Consider the same setup as in Theorem \ref{theorem:random}. When $K\ge 3$, the expected SNR boost achieved by the CSM method in Algorithm \ref{alg:BLD} has a tight order bound:
	\begin{equation}
		\mathbb E[f(\btheta^\text{CSM})] = \frac{\bar\beta^2}{\beta^2_0}\cdot\Theta(N^2)\;\;\text{if}\;\; T = \Omega(N^2(\log N)^3), 
	\end{equation}
	where the expectation is taken over random samples of $\btheta$.
\end{theorem}
\begin{IEEEproof}
	See Appendix \ref{proof:blind}.
\end{IEEEproof}

\begin{figure}[t]
    \centering
    \hspace*{-1.5em}
    \includegraphics[width=4.0cm]{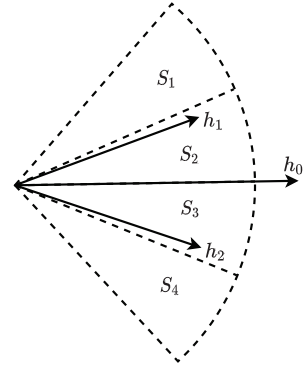}
    \caption{Consider four sectors $\{\mathcal S_1,\mathcal S_2,\mathcal S_3,\mathcal S_4\}$; each sector spans an angle of $\pi/K$. For $N=2$, assume that $h_0$ is located right between $\mathcal S_2$ and $\mathcal S_3$, and that $h_1$ is inside $\mathcal S_2$ but arbitrarily close to $\mathcal S_1$, while $h_2$ and $h_1$ are symmetric about $h_0$. When $K=2$ and $|h_1|=|h_2|$, it follows that $h_1$ and $h_2$ cancel out each other under CPP or CSM, so IRS does not boost SNR.}
    \label{fig:bad_example}
\end{figure}

\begin{remark}
\label{remark:bad_example}
	The main idea of the above proof is to show that $\btheta^\text{CSM}=\btheta^\text{CPP}$ with high probability if $T = \Omega(N^2(\log N)^3)$.
\end{remark}

\begin{remark}
\label{remark:CSM_CPP}
	The condition $K\ge3$ is critical to Theorem \ref{theorem:blind}. When $K=2$, the SNR boost of $\bm\theta^\text{CPP}$ can be arbitrarily close to 1 (i.e., IRS does not bring any gain) in the worst case, and hence $\mathbb E[f(\btheta^\text{CSM})]$ cannot be bounded away from 1. This subtle point is illustrated in Fig.~\ref{fig:bad_example}. An enhanced version of CSM in Section \ref{sec:ECSM} can address this issue.
\end{remark}

\subsection{A Least-Squares Interpretation of CSM}

It turns out that there is an intimate relationship between the proposed blind beamforming method CSM and the least-squares programming. With each $\theta_n$ uniformly and independently drawn in $\Phi_K$, the expectation of the received signal power is given by
\begin{equation}
    \mathbb E[|Y|^2]=\beta_0^2P+\sum^N_{m=1}\beta_m^2P+\sigma^2.
\end{equation}
Moreover, the expectation conditioned on $\theta_n=k\omega$ is given by
\begin{equation}
\label{Rician_power}
	\mathbb E\big[|Y|^2|\theta_n=k\omega\big] = P\big|h_0+h_ne^{jk\omega}\big|^2 + \sum_{m\ne n}\beta_m^2P+\sigma^2.
\end{equation}
With the phase difference $\Delta_n$ defined in \eqref{Delta}, we can write the difference between the above two expectations as a function of $\Delta_n$, that is
\begin{align}
	J_{nk}(\Delta_n) &= \mathbb E[|Y|^2|\theta_n=k\omega]-\mathbb E[|Y|^2]\notag\\
	&= 2\beta_0\beta_nP\cos(k\omega-\Delta_n).
		\label{J:b}
\end{align}
In the meanwhile, we can evaluate each $J_{nk}(\Delta_n)$ empirically as $\widehat J_{nk}$ based on the random samples, i.e.,
\begin{equation}
    \widehat J_{nk} = \frac{1}{|\mathcal Q_{nk}|}\sum_{t\in\mathcal Q_{nk}}|Y_t|^2-\frac{1}{T}\sum^T_{t=1}|Y_t|^2.
\end{equation}
We now estimate $\Delta_n$ by minimizing the gap between $J_{nk}$ and $\widehat J_{nk}$. In particular, with the \emph{square-of-max} metric
\begin{align*}
	\delta_n(\Delta_n)=\left|\max_{1\le k\le K}\big\{J_{nk}(\Delta_n)\big\}-\max_{1\le k\le K}\big\{\widehat J_{nk}\big\}\right|^2,
\end{align*}
the distortion minimizing problem
\begin{subequations}
\label{prob:distortion}
\begin{align}
&\underset{\Delta_1,\ldots,\Delta_N}{\text{minimize}}  \quad \sum^N_{n=1}\delta_n(\Delta_n)\\
&\text{subject to} \quad 0\le\Delta_n<2\pi,\;\;\text{for all}\;\; n=1,\ldots,N
\end{align}
\end{subequations}    
can be optimally solved as
\begin{equation}
\label{est_Delta}
\widehat\Delta_n={k_0}\omega\;\;\text{where}\;\; k_0 = \arg\max_{1\le k\le K} \widehat J_{nk}.
\end{equation}
We then arrive at a somewhat surprising equivalence between $\widehat\Delta_n$ and $\theta^\text{CSM}_n$, as stated in the following proposition.

\begin{proposition}
	The optimal estimate of $\widehat\Delta_n$ in \eqref{est_Delta} is equivalent to the solution $\theta^\text{CSM}_n$ in \eqref{blind:CSM} of the CSM method.
\end{proposition}

The above relationship implies a generic form of CSM. For instance, if we alternatively use the \emph{sum-of-squares} metric
\begin{align*}
	\delta_n(\Delta_n)=\sum^{K}_{k=1}\big|J_{nk}(\Delta_n)-\widehat J_{nk}\big|^2,
\end{align*}
the optimal estimate of $\Delta_n$ in \eqref{prob:distortion} becomes
\begin{equation}
\label{new_Delta_estimation}
{\widehat\Delta}_n=\left\{
\begin{aligned}
&-\arctan\frac{F_n}{E_n}+\frac\pi 2,\;\text{if}\; E_n\ge0,\\
&-\arctan\frac{F_n}{E_n}-\frac\pi 2,\;\text{if}\; E_n<0,\\
\end{aligned}
\right.
\end{equation}
where
\begin{equation}
	{E_n = \sum^{K}_{k=1}}\widehat{J}_{nk}\sin(k\omega)\quad\text{and}\quad F_n = \sum^{K}_{k=1}\widehat{J}_{nk}\cos(k\omega).
\end{equation}
Setting each $\theta^\text{CSM}_n$ to the above $\widehat\Delta_n$ results in a different CSM algorithm.
Actually, we can construct a family of CSM by choosing various metrics $\delta_n(\Delta_n)$. The next subsection further extends CSM to the multiple-antenna and multiple-user cases.



\subsection{Generalized CSM Method}
\label{subsec:MIMO}

Our discussion thus far concentrates on the SNR boost. But we may wish to pursue some other objectives, e.g., the \textcolor{black}{sum rates} across multiple users in a multiple-input multiple-output (MIMO) network. Toward this end, we extend the CSM method to a general utility function. Specifically, we still generate $T$ random samples, and now use a scalar-valued utility $U_t\in\mathbb R$ to quantify the performance of each random sample $\btheta_t$, $t=1,\ldots,T$. For instance, $U_t$ could be set to the \textcolor{black}{sum rates} across multiple users.

We then compute the conditional sample mean of $U_t$ within each subset $\mathcal Q_{nk}$, i.e.,
\begin{equation}
	\widehat{\mathbb E}[U|\theta_n=k\omega] = \frac{1}{|\mathcal Q_{nk}|}\sum_{t\in\mathcal Q_{nk}}U_t.
\end{equation}
Following Algorithm \ref{alg:BLD}, we set each $\theta_n$ to some $\varphi\in\Phi_K$ with the highest conditional sample mean $\widehat{\mathbb E}[U|\theta_n=k\omega]$, i.e.,
\begin{equation}
	\theta^\text{CSM}_{n} = \arg\max_{\varphi\in\Phi_K} \widehat{\mathbb E}[U|\theta_n=\varphi].
\end{equation}
As illustrated in Fig.~\ref{fig:bad_example}, a subtle issue with CSM is that its approximation ratio equals zero when $K=2$. In the next section, we further improve CSM so that its approximation ratio can be bounded away from zero for all $K\ge2$.


\section{Enhanced Conditional Sample Mean Method}
\label{sec:ECSM}

As shown in \cite[Proposition~2]{IRS:opt_21}, the CPP method in \eqref{greedy_alg} attains an approximation ratio of $\cos^2(\pi/K)$, while an enhanced CPP attains a higher approximation ratio of $0.5+0.5\cos(\pi/K)$. Observe that the approximation ratio of the enhanced CPP can be bounded away from zero for any $K\ge2$. Thus far we discuss how to realize CPP without CSI, but now can we also mimic the enhanced CPP blindly?


This section gives a positive answer to the above question. We first sketch the enhanced CPP method in \cite{IRS:opt_21}. Define four sectors on the complex plane as
\begin{multline}
\label{sectors}
\mathcal S_i = \bigg\{u\in\mathbb C:\alpha_0+\frac{(2-i)\omega}{2}\le\mathrm{Arg}(u)\le\alpha_0+\frac{(3-i)\omega}{2}\bigg\},\\
	\text{for}\; i=1,2,3,4,
\end{multline}
which are illustrated in Fig.~\ref{fig:bad_example}. The main idea of the enhanced CPP \cite{IRS:opt_21} is to compare the following three potential solutions: 
\begin{enumerate}[1.]
	\setlength{\itemindent}{1em}
\item  $\btheta'$ that makes each $h_ne^{j\theta_{n}}\in S_1\cup\mathcal S_2$;
\item $\btheta''$ that makes each $h_ne^{j\theta_{n}}\in\mathcal S_2\cup\mathcal S_3$;
\item $\btheta'''$ that makes each $h_ne^{j\theta_{n}}\in\mathcal S_3\cup\mathcal S_4$.
\end{enumerate}
The enhanced CPP in \cite{IRS:opt_21} is basically to compare the above three potential solutions and choose the best. We now modify CSM in Algorithm \ref{alg:BLD} to recover the enhanced CPP. First of all, $\btheta''$ can be recognized as $\bm\theta^\text{CPP}$, which can be blindly recovered by CSM as stated in Remark \ref{remark:bad_example}, so it remains to recover the other two potential solutions $\bm\theta'$ and $\bm\theta'''$.

Suppose that we already obtain $\bm\theta''$, so each $h_ne^{j\theta_n}$ belongs to either $\mathcal S_2$ or $\mathcal S_3$. If we rotate all those $h_ne^{j\theta_n}$ in $\mathcal S_3$ counterclockwise by an angle of $\omega$, then each $h_ne^{j\theta_n}$ belongs to either $\mathcal S_1$ or $\mathcal S_2$ afterwards, i.e., the above procedure transforms $\bm\theta''$ into 
$\bm\theta'$.
Now the key is to decide which $h_ne^{j\theta_n}$ belong to $\mathcal S_3$ under $\btheta''$. 
Let us assume that $K\ge3$; the case of $K=2$ will be dealt with in Remark \ref{ECSM:K2}. Consider a particular channel $h_ne^{j\theta''_n}$ that currently belongs to $\mathcal S_3$ under $\theta''_n$. If we rotate it \emph{counterclockwise} by an angle of $\omega$, the new channel $h_ne^{j(\theta''_n+\omega)}$ would belong to $\mathcal S_1$, and it can be shown that 
$\big|\mathrm{Arg}(h_0)-\mathrm{Arg}(h_ne^{j(\theta''_n+\omega)})\big|\le \omega$.
Alternatively, if $h_ne^{j\theta''_n}$ is rotated \emph{clockwise} by an angle of $\omega$, then
we must have 
$
\big|\mathrm{Arg}(h_0)-\mathrm{Arg}(h_ne^{j(\theta''_n-\omega)})\big|\ge \omega$.
Furthermore, because $\mathbb E\big[|Y|^2|\theta_n=k\omega\big]$ increases monotonically when the angle between $h_n$ and $h_ne^{jk\omega}$ becomes smaller as observed from \eqref{Rician_power}, for this particular $h_ne^{j\theta''_n}\in\mathcal S_3$, we must have $\mathbb E[|Y|^2|\theta_n=k\omega+\omega]\ge\mathbb E[|Y|^2|\theta_n=k\omega-\omega]$. Likewise, we must have $\mathbb E[|Y|^2|\theta_n=k\omega+\omega]\le\mathbb E[|Y|^2|\theta_n=k\omega-\omega]$ whenever
$h_ne^{j\theta''_n}\in\mathcal S_2$. As a result, we can figure out which sector each $h_ne^{j\theta''_n}$ belongs to by comparing $\mathbb E[|Y|^2|\theta_n=k\omega+\omega]$ and $\mathbb E[|Y|^2|\theta_n=k\omega-\omega]$.

In practice, we evaluate $\mathbb E[|Y|^2|\theta_n=k\omega]$ empirically as $\widehat{\mathbb E}[|Y|^2|\theta_n=k\omega]$ in (\ref{conditional_mean}), so $\btheta'$ can be obtained from $\btheta''$ as
\begin{equation}
	\label{psi:12}
	\theta'_n = \theta''_{n} + \omega\Lambda_{n},\;\forall n=1,\ldots,N,
\end{equation}
where
\begin{equation}
\label{delta}
	\Lambda_n= \mathbbm{1}_{+}\Big(\widehat{\mathbb E}[|Y|^2|\theta_n=\theta''_n+\omega]-\widehat{\mathbb E}[|Y|^2|\theta_n=\theta''_n-\omega]\Big)
\end{equation}
with the indicator function $\mathbbm{1}_{+}(x)=1$ if $x\ge0$ and $\mathbbm{1}_{+}(x)=0$ otherwise.

Furthermore, notice that $\btheta'''$ can be obtained from $\btheta'$ by rotating every $h_ne^{j\theta'_n}$ clockwise by an angle of $\omega$, i.e.,
\begin{equation}
	\label{psi:34}
	\theta'''_n = \theta'_{n} - \omega,\;\;\text{for all}\;\; n=1,\ldots,N.
\end{equation}
Following the approximation algorithm in \cite{IRS:opt_21}, we choose the best out of the three potential solutions:
\begin{equation}
	\label{ECSM}
	\btheta^{\text{ECSM}}=\arg\max_{\bm\vartheta\in\{\btheta',\btheta'',\btheta'''\}} \widehat{\mathbb E}\big[|Y|^2|\btheta=\bm\vartheta\big].
\end{equation}
The above method is referred to as the \emph{Enhanced Conditional Sample Mean (ECSM)}.
Algorithm \ref{alg:E-BLD} summarizes it.

\begin{figure*}[t]
	\centering
	\centerline{\includegraphics[width=15cm]{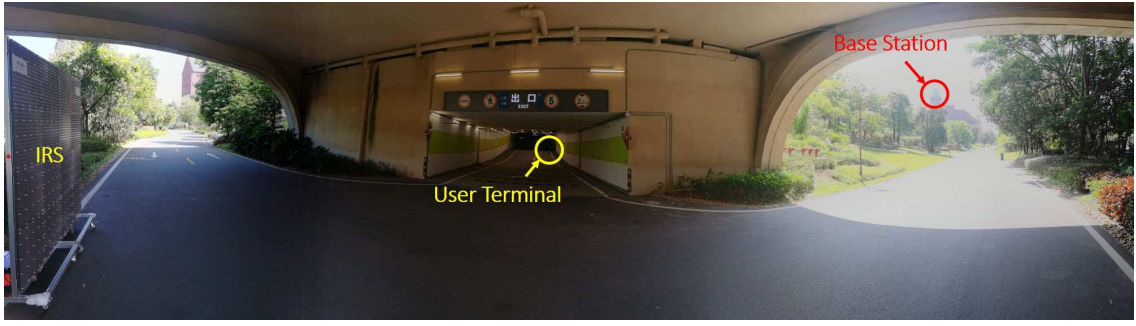}}
	\caption{A panoramic view of the field test site. The base station is located on a 20-meter-high terrace while the user terminal is located inside an underground parking lot. The IRS is placed at the entrance of the parking lot. The IRS is approximately 250 meters away from the base station, and the user terminals are approximately 40 meters away from the IRS.} 
	\label{fig:panorama}
\end{figure*}

\begin{algorithm}[t]
\caption{Enhanced Conditional Sample Mean (ECSM)}
\label{alg:E-BLD}
\begin{algorithmic}[1]
\State\textbf{input:} $\Phi_K$, $N$, $T$;
\State run Algorithm \ref{alg:BLD} and let $\btheta''=\btheta^\text{CSM}$;
\State compute $\btheta'$ by
(\ref{psi:12}) and compute $\btheta'''$ by (\ref{psi:34});
\For{$n=1,2,\ldots,N$}
	\State compute $\hat{\mathbb E}[|Y_t|^2|\theta_n=\vartheta]$, for each $\vartheta\in\{\btheta',\btheta'',\btheta'''\}$;
	\State decide $\theta^\text{ECSM}_n$ according to (\ref{ECSM});
\EndFor
\State\textbf{output:} $\btheta^\text{ECSM}=(\theta^\text{ECSM}_1,\ldots,\theta^\text{ECSM}_N)$.
\end{algorithmic}
\end{algorithm}

\begin{remark}
	The way we compute $\btheta'$ and $\btheta'''$ based on $\btheta''$ according to (\ref{psi:12})--(\ref{psi:34}) is of practical significance. It only requires measuring the received signal power $|Y_t|^2$ rather than the in-phase and quadrature branches of the received signal.
\end{remark}
	
\begin{remark}
\label{ECSM:K2}
	When $K=2$, we can no longer use (\ref{psi:12})--(\ref{psi:34}) to obtain $\btheta'$ and $\btheta'''$ because the clockwise rotation by an angle of $\omega$ is now identical to the counterclockwise. In this case, we send the same symbol $X_0$ across the random samples and redefine $\Lambda_n$ in \eqref{psi:12} as $\Lambda_n= \mathbbm{1}_{+}\big(\mathrm{Im}\big\{\widehat{\mathbb E}[Y|\theta_n=\theta''_n]\big\}\big)$.
\end{remark}

As compared to CSM, the main advantage of ECSM is that it achieves the upper bound in Proposition \ref{prop:upper_bound} for any $K\ge2$, as specified in the following theorem.
\begin{theorem}
\label{theorem:enhanced_blind}
Consider the same setup as in Theorem \ref{theorem:random}. 
For any $K\ge2$, the expected SNR boost achieved by the
ECSM method in Algorithm \ref{alg:E-BLD} has a tight order bound:
\begin{equation}
	\mathbb E[f(\btheta^\text{ECSM})] = \frac{\bar\beta^2}{\beta^2_0}\cdot \Theta(N^2)\;\;\text{if}\;\; T = \Omega(N^2(\log N)^3), 
\end{equation}
where the expectation is taken over random samples of $\btheta$. In particular, the condition can be relaxed to $T = \Omega(N^2\log N)$ when $K=2$.
\end{theorem}
\begin{IEEEproof}
	See Appendix \ref{proof:ECSM}.
\end{IEEEproof}

We conclude this section by emphasizing the computational efficiency of ECSM; it turns out that ECSM and CSM both have a computational complexity of $O(TN)$.



\begin{figure}[t]
    \centering
    \includegraphics[width=8cm]{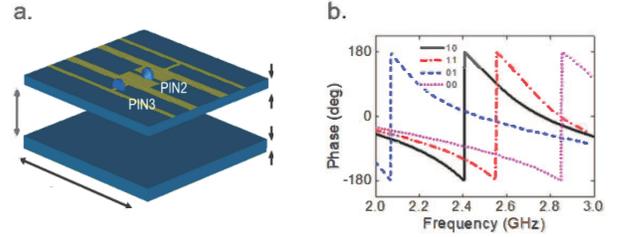}
    \caption{The {\footnotesize{ON-OFF}} state of a PIN diode results in two distinct resonance frequencies in the series RLC circuit, which correspond to two phase shifts. Further, with a pair of PIN diodes integrated into each reflective element, we can realize
	four phase shifts by controlling the respective {\footnotesize{ON-OFF}} states of the two PIN diodes.}
    \label{fig:prototype}
\end{figure}

\section{Experiments}
\label{sec:experiments}

This section comprises two parts. First, we validate the performance of the proposed algorithms CSM and ECSM in a real-world 5G network as shown in Fig.~\ref{fig:panorama}. Second, we conduct simulations to compare blind beamforming with the conventional CSI based methods.

\subsection{Field Tests}
\label{subsec:field_tests}

The field tests are carried out for downlink transmission from a public base station to a user terminal in bandwidth 200 MHz at 2.6 GHz. Throughout our tests, \emph{zero} knowledge about the base station side is assumed, and \emph{zero} coordination from the base station side is required. In other words, the IRS works in a plug-and-play fashion, and its deployment is completely invisible to service provider.

\begin{figure}[t]
    \centering
    \includegraphics[width=7cm]{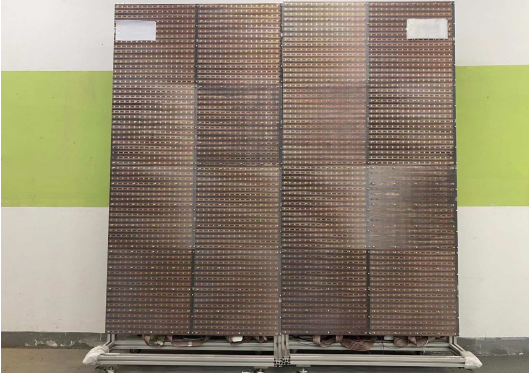}
    \caption{\footnotesize{The IRS is formed by a $4\times4$ array of reflecting tiles. Each reflecting tile is 50cm$\times$50cm large and consists of 16 reflective elements.}}
    \label{fig:IRS}
\end{figure}

The hardware realization of each reflective element is illustrated in Fig.~\ref{fig:prototype}. Moreover, as shown in Fig.~\ref{fig:IRS}, the IRS is formed by $16$ ``reflecting tiles''---a tiny IRS prototype that is $50\,$cm$\,\times\,50\,$cm large---arranged in a $4\times4$ array. Each reflecting tile consists of $16$ reflective elements, so the assembled large IRS consists of $256$ reflective elements in total. There are $4$ phase shift choices $\{0,\pi/2,\pi,3\pi/2\}$ for each individual reflective element.

\begin{figure}[t]
    \centering
    \centerline{\includegraphics[width=7cm]{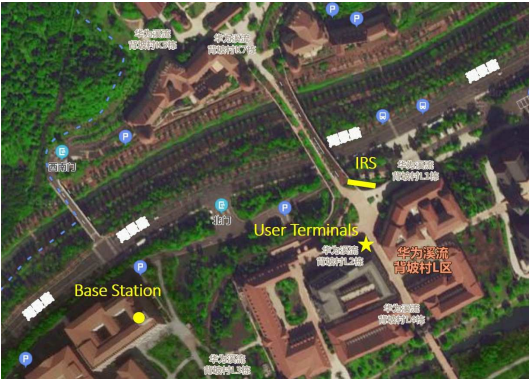}}
    \caption{\footnotesize{A satellite image of the field test site. The base station and the IRS are outdoor while the user terminals are indoor.}}
    \label{fig:map}
\end{figure}

\begin{figure}[t]
    \centering
    \includegraphics[width=7cm]{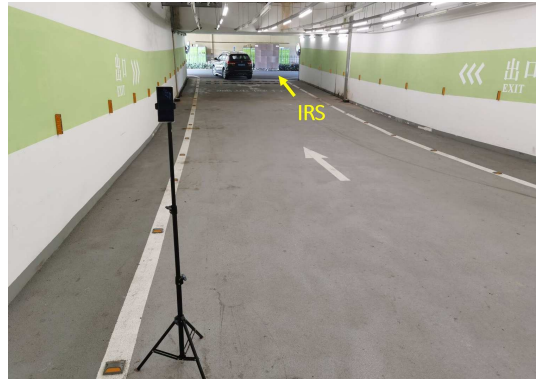}
    \caption{\footnotesize{The view from the user terminal toward the IRS.}}
    \label{fig:UE}
\end{figure}


As shown in Fig.~\ref{fig:panorama} and Fig.~\ref{fig:map}, the base station is located on a 20-meter-high terrace while the user terminal is located in an underground parking lot. There is no line-of-sight propagation from the base station to the user terminal. The IRS is placed outdoors near the entrance of the parking lot. The distance from the base station to the IRS is approximately 250 meters; the distance from the IRS to the user terminal is approximately 40 meters. It is worth remarking that the wireless environment is highly volatile in our case because of the busy traffic in the parking lot, as can be observed from Fig.~\ref{fig:UE}.



The random-max sampling (RMS) algorithm uses a total of $2560$ random samples, i.e., $T=10N$. We compare RMS with the conditional sample mean (CSM) algorithm and the enhanced conditional sample mean (ECSM) algorithm. In addition, we include in the tests a baseline method called OFF, which simply deactivates all the reflective elements so that the IRS reduces to a metal surface without beamforming. 

\begin{figure}[t]
\centering
\hspace*{-1.5em}
\includegraphics[width=9cm]{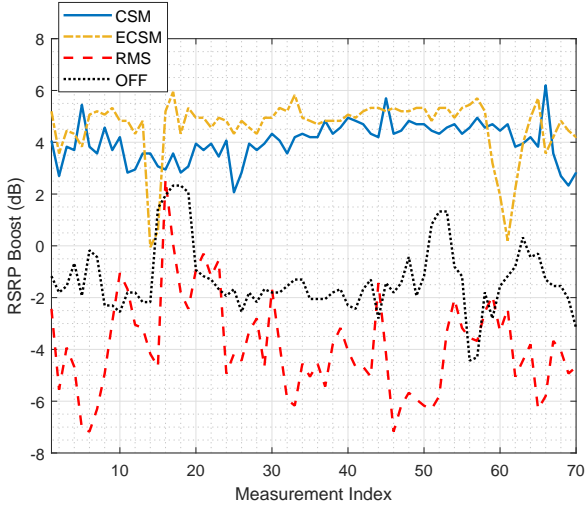}
\caption{RSRP boost for SISO transmission.}
\label{fig:RSRP}
\end{figure}

We start with the single-input single-output (SISO) transmission, aiming to improve the SNR boost. There are two measurements: Reference Signal Received Power (RSRP) and Signal-to-Interference-plus-Noise Ratio (SINR). Notice that the SNR cannot be measured directly because of co-channel interference. Following the definition of the SNR boost, we let the RSRP (or SINR) boost be the ratio between the achieved RSRP (or SINR) and the baseline RSRP (or SINR) without IRS. Fig.~\ref{fig:RSRP} shows the RSRP boosts achieved by the various methods. It can be seen that CSM and ECSM outperform the other methods significantly. For instance, there is an approximately 5 dB gap between ECSM and OFF. As shown in the figure, although ECSM encounters two sharp drops, which are due to the shadowing effect caused by vehicles, its overall performance is still more consistent over time than RMS and OFF. Observe from Fig.~\ref{fig:RSRP} that the RSRP boost by OFF is mostly below $0$ dB; the reason is that the reflected signals without proper beamforming can result in a destructive superposition. Observe also that RMS yields the worst RSRP performance, even $4$ dB lower than not using IRS. This surprising result indicates that, in a complicated wireless environment with interference and noise, the beamformer decision based on the best single sample is not reliable.

\begin{figure}[t]
\centering
\hspace*{-1.5em}
\includegraphics[width=9cm]{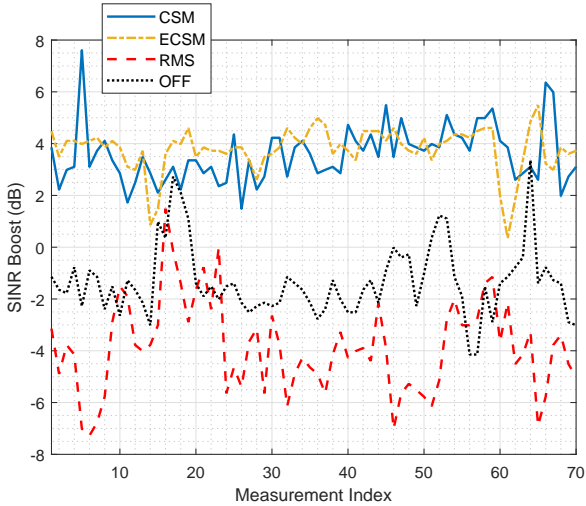}
\caption{SINR boost for SISO transmission.}
\label{fig:SINR}
\vspace{-1em}
\end{figure}

We further compare the SINR boosts of the various methods in Fig.~\ref{fig:SINR}. It can be seen that the SINR boosts and the RSRP boosts have similar profiles. The average RSRP boosts and the average SINR boosts are summarized in Table \ref{tab:RSRP}. According to the table, the SINR gain is smaller than the RSRP gain. One reason for this gain reduction is that IRS incurs additional reflected interference. Nevertheless, the constructive effect on the desired signals outweighs that on the interfering signals. As a result, CSM and ECSM can still bring considerable performance gains as compared to the benchmark methods and not using CSI.

\begin{figure}[t]
\centering
\hspace*{-1.5em}
\includegraphics[width=9cm]{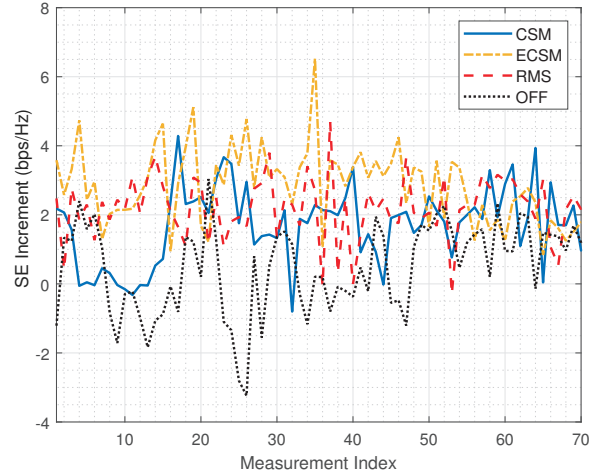}
\caption{SE increment for MIMO transmission.}
\label{fig:SE}
\end{figure}

Moreover, we consider the MIMO transmission. In our case, the base station has $64$ transmit antennas while the user terminal has $4$ receive antennas, so at most $4$ data streams are supported. Because the base station is a black box to us, how the transmit precoding is performed is unknown. As stated in Section \ref{subsec:MIMO}, it is difficult to define a scalar-valued SNR boost in this scenario. We adopt the generalized CSM and ECSM in Section \ref{subsec:MIMO} with the \emph{Spectral Efficiency (SE)} utility. Thus, we measure the SE in bps/Hz at the user terminal for each random sample. Fig.~\ref{fig:SE} shows the SE increments by the various algorithms against the baseline SE without IRS. Observe that all the algorithms can bring improvements, although OFF occasionally gives negative effects. The figure shows that RMS becomes more robust in the MIMO case. Actually, RMS is sometimes even better than CSM and ECSM, but it still has inferior performance on average. The average SE increment results summarized in Table \ref{tab:RSRP} agree with what we observe from Fig.~\ref{fig:SE}.

\renewcommand{\arraystretch}{1.0}
\begin{table*}[t]
\renewcommand{\arraystretch}{1.3}
\footnotesize
\centering
\caption{\small Average Performance of the Various Algorithms}
\begin{tabular}{|c||c|c||c|}
\hline
& \multicolumn{2}{c||}{SISO} & MIMO \\
\hline
Algorithm & RSRP Boost (dB) & SINR Boost (dB) & SE Increment (bps/Hz) \\
\hline
\hline
CSM & $4.02$ & $3.57$ & $2.02$ \\
\hline
ECSM & $4.62$ & $3.81$ & $2.08$ \\
\hline
RMS & $-3.93$ & $-3.84$ & $1.97$ \\
\hline
OFF & $-1.69$ & $-1.69$ & $0.77$ \\
\hline
\end{tabular}
\label{tab:RSRP}
\vspace{-1em}
\end{table*}

\subsection{Simulation Test: Blind Beamforming vs. CSI Approach}
\label{sec:simulations}

As remarked earlier, it is difficult to perform channel estimation in an IRS system not only because the reflected channels are comparatively weak but also because it is incompatible with the existing 5G protocol. Thus, we would compare the CSI approach with the proposed blind beamforming schemes via simulations. The channel model follows the previous works \cite{rui_SDP_BF_19,Yuen_DBF_20,wy_AI_21}. The background channel $h_0$ is modeled as
\begin{equation}
    \textcolor{black}{h_0 = 10^{-\mathsf{PL}_0/20}\phi_0,}
\end{equation}
where the transmitter-to-receiver pathloss $\mathsf{PL}_0$ (in dB) is computed as $\mathsf{PL}_0=32.6+36.7\log_{10}(d_0)$, with $d_0$ in meters denoting the distance between the transmitter and the receiver, while the Rayleigh fading component $\phi_0$ is drawn from the Gaussian distribution $\mathcal{CN}(0,1)$. The cascaded reflected channel $h_n$ 
is modeled as
\begin{equation}
    \textcolor{black}{h_n = 10^{-(\mathsf{PL}_1+\mathsf{PL}_2)/20}\phi_{n1}\phi_{n2},\;\; n=1,\ldots,N,}
\end{equation}
where $\mathsf{PL}_1$ and $\mathsf{PL}_2$ are both based on the pathloss model $\mathsf{PL}_0=30+22\log_{10}(d)$, with $d$ in meters respectively denoting the transmitter-to-IRS distance and the IRS-to-receiver distance, {while the Rayleigh fading components $\phi_{n1}$ and $\phi_{n2}$ are drawn from the Gaussian distribution $\mathcal{CN}(0,1)$ independently across $n=1,\ldots,N$.} We use the following parameters unless otherwise stated. The transmit power level $P$ equals $30$ dBm. The background noise power $\sigma^2=-90$ dBm unless otherwise stated. The locations of the transmitter, IRS, and receiver are respectively denoted by the 3-dimensional coordinate vectors $(50,-200,20)$, $(-2,-1,0)$, and $(0,0,0)$ in meters. Unless otherwise stated, the number of reflective elements $N$ and the number of random samples $T$ are both set to $500$.

\begin{figure}[t]
\begin{minipage}[b]{1.0\linewidth}
		\centering
		\centerline{\includegraphics[width=9.5cm]{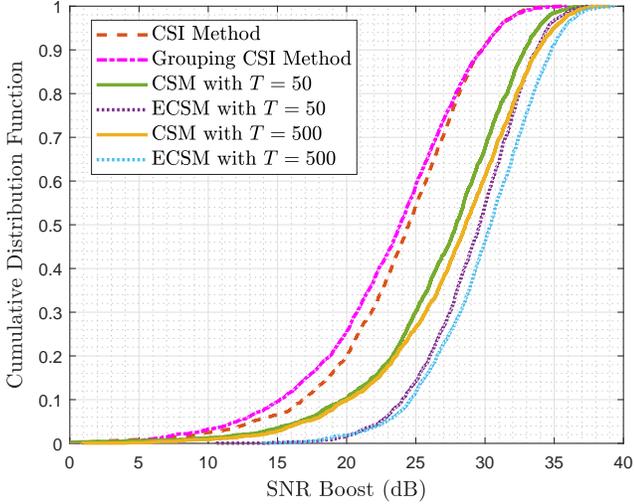}}
\end{minipage}
\caption{Cumulative distribution of SNR boost when $\sigma^2=-90$ dBm.}
\label{fig:CDF_SISO_SU_90}
\end{figure}

\begin{figure}[t]
\vspace{-1em}
\begin{minipage}[b]{1.0\linewidth}
	  \centering
	  \centerline{\includegraphics[width=9.5cm]{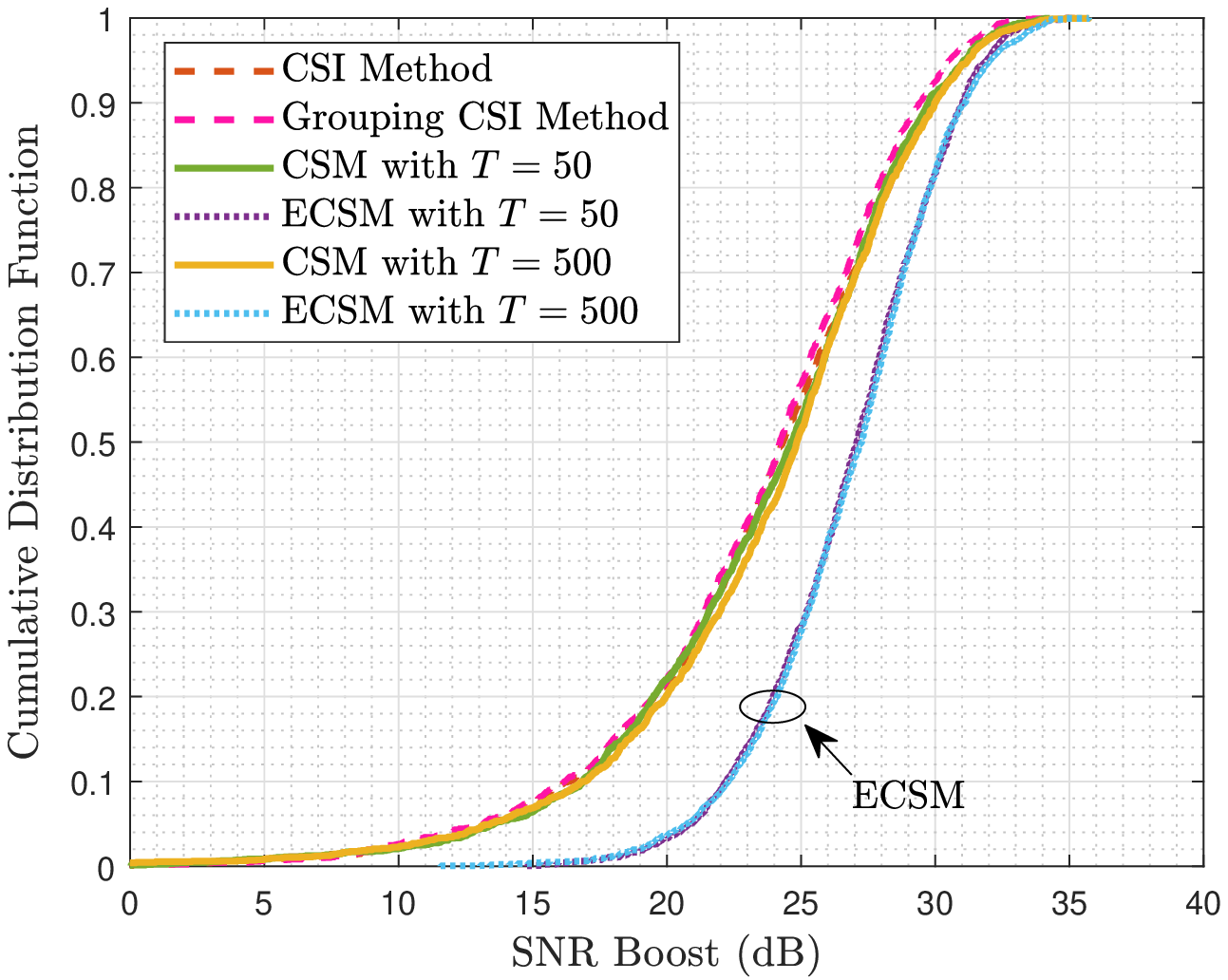}}
\end{minipage}
\caption{Cumulative distribution of SNR boost when $\sigma^2=-50$ dBm.}
\label{fig:CDF_SISO_SU_50}
\end{figure}

\begin{figure}[t]
\begin{minipage}[b]{1.0\linewidth}
	  \centering
	  \centerline{\includegraphics[width=9.5cm]{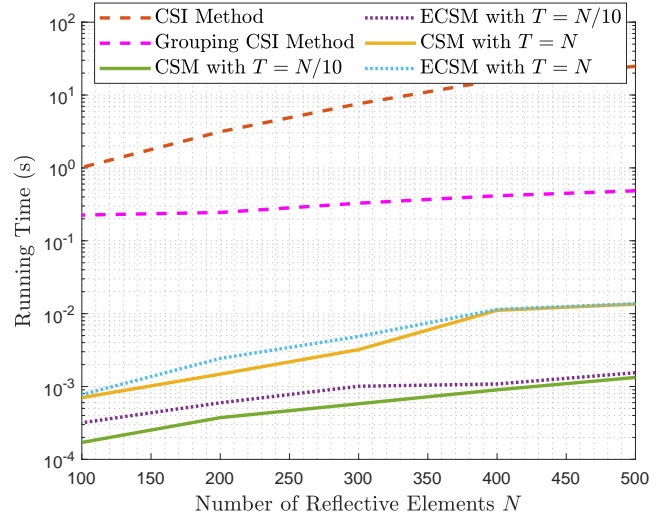}}
\end{minipage}
\caption{Running time versus number of reflective elements $N$.}
\vspace{-1em}
\label{fig:Aver_SISO_SU_time}
\end{figure}

\begin{figure}[t]
\begin{minipage}[b]{1.0\linewidth}
	  \centering
	  \centerline{\includegraphics[width=9.5cm]{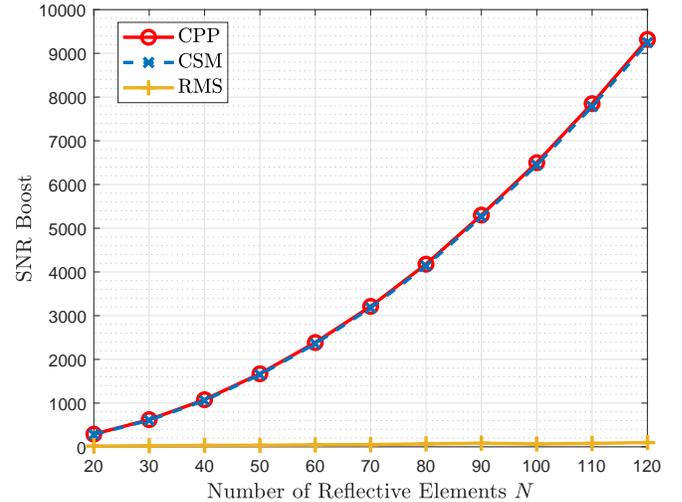}}
\end{minipage}
\caption{SNR boost versus number of reflective elements $N$.}
\label{fig:scaling}
\end{figure}

\begin{figure}[t]
\begin{minipage}[b]{1.0\linewidth}
	  \centering
	  \centerline{\includegraphics[width=9.5cm]{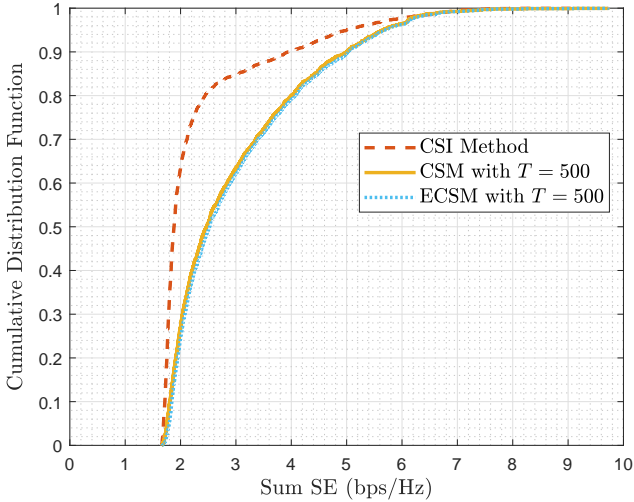}}
\end{minipage}
\caption{Cumulative distribution of sum SE when $\sigma^2=-90$ dBm.}
\label{fig:CDF_MISO_MU_90}
\end{figure}

\begin{figure}[t]
\begin{minipage}[b]{1.0\linewidth}
	  \centering
	  \centerline{\includegraphics[width=9.5cm]{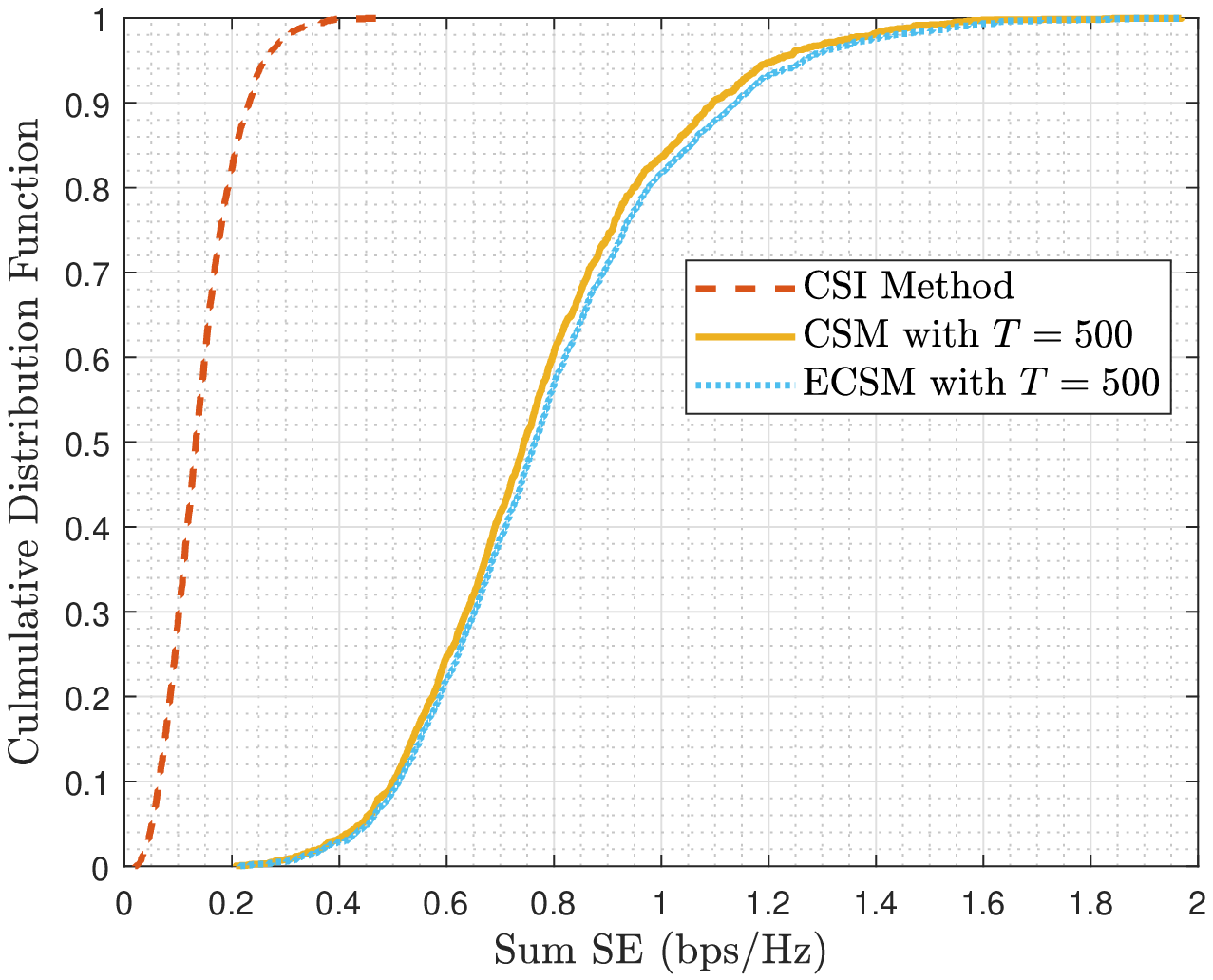}}
\end{minipage}
\caption{Cumulative distribution of sum SE and $\sigma^2=-50$ dBm.}
\label{fig:CDF_MISO_MU_50}
\end{figure}

Aside from the proposed CSM and ECSM, the following two baseline methods are included in the comparison:
\begin{itemize}
    \item \emph{CSI Method:} First use the DFT method to estimate channels and then use the SDR method (for the single-user case) or the stochastic successive convex approximation (for the multi-user case) to optimize $\bm\theta$.

    \item \emph{Grouping CSI Method:} Improve the efficiency of channel estimation in the above CSI method by grouping the reflective elements as suggested in \cite{ZhangRui_ICC_2020,grouping_Zhangrui_LWC_2020}. We divide the reflective elements equally into $50$ groups.
\end{itemize}

Fig.~\ref{fig:CDF_SISO_SU_90} compares the cumulative distribution functions of the different algorithms for a single link when the background noise power $\sigma^2=-90$ dBm. It can be seen that ECSM outperforms CSM while CSM outperforms CSI method given the same training overhead $T$. (ECSM and CSM only require the receiver to measure the received signal power whereas CSI method requires the signal phase in addition.) It can be also seen that grouping CSI method is inferior to CSI method even though it estimates channel more efficiently. We remark that reducing the number of samples to $50$ does not impact the performance of CSM and ECSM much. When the noise power $\sigma^2$ is raised to $-50$ dBm, as shown in Fig.~\ref{fig:CDF_SISO_SU_50}, the CSI method and CSM achieve similar performance. In this case, the advantage of ECSM over the other algorithms becomes greater.

We further compare the running time of the different algorithms. First, for the current prototype, we can take up to $10$ samples per second, so it takes less than one minute in total to obtain $500$ samples; the sampling time can be further cut down to $5$ seconds if we let $T=50$ which still yields good performance as formerly shown in Fig.~\ref{fig:CDF_SISO_SU_90} and Fig.~\ref{fig:CDF_SISO_SU_50}. Second, after sampling, the data processing time is shown in Fig.~\ref{fig:Aver_SISO_SU_time}. Observe that CSM and ECSM require much less time than the two CSI methods. Moreover, Fig.~\ref{fig:scaling} shows how the SNR boost scales with the number of reflective elements. We set $T= \lceil(N^2(\log N)^3)\rceil$. According to the figure, CPP and CSM yield similar performance with a quadratic scaling rate of SNR boost, whereas the SNR boost of RMS grows linearly and slowly in $N$. This result agrees with Theorems \ref{theorem:random} and \ref{theorem:blind}.

Finally, we consider maximizing the sum SE across multiple receivers. Add three more single-antenna receivers to the network, which are located at $(0,1,0)$, $(1,0,0)$ and $(1,1,0)$, respectively. The transmitter is now deployed with 4 transmit antennas; it serves the four receivers via spatial multiplexing. The performance of the optimized passive beamforming is averaged over the randomly generated active beamforming at the transmitter. To deal with multiple users, we take the utility function $U_t$ in Section \ref{subsec:MIMO} to be the sum SE of each sample.
Fig.~\ref{fig:CDF_MISO_MU_90} compares the different methods when $\sigma^2=-90$ \textcolor{black}{dBm}. According to the figure, CSM and ECSM are close to each other; CSI has worse performance but is close to the other algorithms in the low sum SE regime. Hence, if channels can be estimated accurately, the model-driven approach like the two-timescale method \cite{mm_2timescale_BF_21} can achieve good performance. However, in the low-SNR regime as shown in Fig.~\ref{fig:CDF_MISO_MU_50} with $\sigma^2=-50$ dBm, the CSI method becomes significantly worse than the other methods. But CSM and ECSM still have similar performance. A possible reason is that simply letting sum SE be the utility function is suboptimal, so a more sophisticated choice of $U_t$ for ECSM could potentially yield much higher boost.



\section{Conclusion}
\label{sec:conclusion}

This paper proposes a statistical approach to the IRS beamforming that is fundamentally different from most of the existing methods in that it does not involve any channel estimation. The proposed algorithm only requires a polynomial number of random samples to provide the highest possible SNR boost---which is quadratic in the number of reflective elements. In contrast, the standard method of random-max sampling can only give a linear boost. Our field experiments show that the proposed algorithm can achieve significant performance gains consistently over time even in a highly volatile wireless environment.



\appendices

\section{Proof of Theorem \ref{theorem:random}}
\label{proof:RMS}

\subsection{Lower Order Bound in (\ref{theorem:random:lower})}
\label{proof:RMS:lower}

We first introduce a multidimensional version of the Berry-Ess\'{e}en theorem for non-i.i.d. random vectors.
\begin{theorem}[Theorem 1.1 in \cite{Bentkus_theory_prob_appl05}]
\label{theorem:BE}
Consider $N$ independent (but not necessarily identically distributed) random vectors $\bx_1,\ldots,\bx_N\in\mathbb R^d$ with zero mean $\mathbb E[\bx_n]=\bm0$, $n=1,\ldots,N$. Write $\bs=\sum^N_{n=1}\bx_n$, $\bW^2 = \mathbb E[\bs\bs^{\mathsf{T}}]$, $\bW$ the positive root of $\bW^2$ (assuming that $\bW^2\succ\mathbf0$),  $\xi_n=\mathbb E[\|\bW^{-1}\bx_n\|^3]$, and $\xi=\sum^N_{n=1}\xi_n$. Letting $\mathcal C$ be the class of all convex subsets of $\mathbb R^d$, there exists an absolute positive constant $\rho>0$ such that
\begin{equation}
	\sup_{\mathcal A\in\mathcal C}\Big|\mathbb{P}\{\bs\in\mathcal A\}-\mathbb{P}\{\bu\in\mathcal A\}\Big|\le  \sqrt[4]{d}\rho\xi,
\end{equation}
where $\bu$ is a $d$-dimensional Gaussian vector with $\mathbb E[\bu]=\bm0$ and $\mathbb E[\bu\bu^{\mathsf{T}}]=\bW^2$.
\end{theorem}

The following corollary specializes the above theorem to our problem case.

\begin{corollary}
\label{corollary:1}
If every phase shift $\theta_n$ is drawn from the set $\Phi_K$ uniformly and independently, there exists an absolute positive constant $\rho>0$ such that the complementary cumulative distribution function (CCDF) of the received signal power $|Y|^2$ is bounded as
\begin{equation}
\label{lemma:ccdf:bound}
	\Big|\mathbb{P}\big\{|Y|^2\ge\gamma\big\}-e^{-\lambda\gamma}\Big|\le\sqrt[4]{2}\rho\xi
\end{equation}
given any threshold $\gamma>0$, where 
\begin{equation}
\label{lambda}	
	\lambda = \left(\sum^N_{n=0}\beta_n^2P+\sigma^2\right)^{-1}
	=\left(h_0P+N\bar\beta^2P+\sigma^2\right)^{-1}
\end{equation}
and
\begin{equation}
\label{xi}
	\xi = (2\lambda)^{3/2}\Bigg(\sum^N_{n=0}\beta^3_nP^{3/2}+\sigma^3\Bigg)\cdot\mathbb E[|u|^3]
\end{equation}
with a unit complex Gaussian random variable $u\sim\mathcal{CN}(0,1)$.
\end{corollary}
\begin{IEEEproof}
The main idea is to rewrite each complex number as a real 2-dimensional vector in $\mathbb R^2$. Recall that $Y=h_0X+\sum^N_{n=1}h_ne^{j\theta_n}X+Z$. Let $\bx_0=(\mathrm{Re}\{h_0X\},\mathrm{Im}\{h_0X\})^\mathsf{T}$, $\bx_n=(\mathrm{Re}\{h_ne^{j\theta_n}X\},\mathrm{Im}\{h_ne^{j\theta_n}X\})^\mathsf{T}$ for $n=1,\ldots,N$, and $\bx_{N+1}=(\mathrm{Re}\{Z\},\mathrm{Im}\{Z\})^\mathsf{T}$. Notice that $\bx_0,\ldots,\bx_{N+1}$ are zero-mean independent random vectors, so Theorem \ref{theorem:BE} applies. With $\bs=\sum^{N+1}_{n=0}\bx_n$, we compute $\bW^2$ as $\mathbb E[\bs\bs^\mathsf{T}]=\frac{1}{2}(\sum^N_{n=0}\beta^2_nP+\sigma^2)\bI$, and thus $\bW=\frac{1}{\sqrt2}(\sum^N_{n=0}\beta^2_nP+\sigma^2)^{\frac12}\bI$. We then obtain $\xi$ as in (\ref{xi}).

Consider a convex subset
\begin{equation}
	\mathcal A = \big\{\bv\in\mathbb R^2:\|\bv\|^2<\gamma\big\}.
\end{equation}
For a Gaussian random variable $\bu\sim\mathcal{N}(\mathbf0,\bW^2)$, we have $\mathbb P\{\bu\in\mathcal A\}=1-e^{-\lambda\gamma}$ with $\lambda=(\sum^N_{n=0}\beta_n^2P+\sigma^2)^{-1}$. Applying Theorem \ref{theorem:BE}, we arrive at
\begin{equation}
	\label{rho_mu_bound:1}
		\Big|\mathbb{P}\big\{\bs\in\mathcal A\big\}-\mathbb{P}\big\{\bu\in\mathcal A\big\}\Big|\le \sqrt[4]{2}\rho\xi,
\end{equation}
which can be rewritten as
\begin{equation}
	\label{rho_mu_bound:2}
		\Big|\mathbb{P}\big\{\bs\notin\mathcal A\big\}-\mathbb{P}\big\{\bu\notin\mathcal A\big\}\Big|\le \sqrt[4]{2}\rho\xi
\end{equation}	
since $\mathbb{P}\big\{\bs\in\mathcal A\big\}=1-\mathbb{P}\big\{\bs\notin\mathcal A\big\}$ and $\mathbb{P}\big\{\bu\in\mathcal A\big\}=1-\mathbb{P}\big\{\bu\notin\mathcal A\big\}$. 
Substituting $\mathbb P\{\bu\notin\mathcal A\}=e^{-\lambda\gamma}$ and $\mathbb{P}\big\{\bs\notin\mathcal A\big\}=\mathbb{P}\big\{|Y|^2\ge\gamma\big\}$ in (\ref{rho_mu_bound:2}) gives the bound in (\ref{lemma:ccdf:bound}).
\end{IEEEproof}

The following lemma characterizes the behavior of the noise power $|Z_t|^2$ under random-max sampling.
\begin{lemma}
\label{lemma:2}
	Given a sequence of i.i.d. noise $Z_t\sim\mathcal{CN}(0,\sigma^2)$, $t=1,\ldots,T$, we have
	\begin{equation}
	\label{Zt:Theta}
		\mathbb E\bigg[\max_{1\le t\le T}|Z_t|^2\bigg]=\sigma^2\cdot\Theta(\log T).
	\end{equation}
\end{lemma}
\begin{IEEEproof}
	Each $|Z_t|^2$ has an exponential distribution with the scale parameter $\sigma^2$, so 
	\begin{equation}
		\mathbb P\big\{|Z_t|^2\le\tau\big\}=1-e^{-\tau/\sigma^2}
	\end{equation}
	and further
	\begin{align}
		\mathbb P\bigg\{\max_{1\le t\le T}|Z_t|^2>\tau\bigg\}
		&= 1 - (1-e^{-\tau/\sigma^2})^T.
	\end{align}
	With $\tau=\sigma^2\log T$, we use Markov's inequality to show that
	\begin{align}
	\label{Zt:Omega}
		\mathbb E\bigg[\max_{1\le t\le T}|Z_t|^2\bigg]&\ge \tau\cdot \mathbb P\bigg\{\max_{1\le t\le T}|Z_t|^2>\tau\bigg\}\notag\\
		&=\sigma^2\log T\cdot\big(1-(1-1/T)^T\big)\notag\\
		&\ge \sigma^2\log T\notag\cdot(1-e^{-1})\\
		&=\sigma^2\cdot\Omega(\log T).
	\end{align}
	Moreover, by Jensen's inequality, we have
	\begin{align}
		\exp\bigg(\frac{1}{2\sigma^2}\mathbb E\bigg[\max_{1\le t\le T}|Z_t|^2\bigg]\bigg)
		&\le \mathbb E\bigg[\exp\bigg(\frac{1}{2\sigma^2}\cdot\max_{1\le t\le T}|Z_t|^2\bigg)\bigg]\notag\\
		&\le\sum^T_{t=1}\mathbb E\bigg[\exp\bigg(\frac{1}{2\sigma^2}|Z_t|^2\bigg)\bigg]\notag\\
		&=2T.
		\label{Z:jensen}
	\end{align}
	Taking the logarithm of both sides yields
	\begin{align}
		\label{Zt:O}
		\mathbb E\bigg[\max_{1\le t\le T}|Z_t|^2\bigg]&\le 2\sigma^2\log(2T)=\sigma^2\cdot O(\log T).
	\end{align}
	Combining (\ref{Zt:Omega}) and (\ref{Zt:O}) gives the order bound in (\ref{Zt:Theta}). 
\end{IEEEproof}

In light of Corollary \ref{corollary:1} and Lemma \ref{lemma:2}, we now turn to showing that $\mathbb E[f(\btheta^\text{RMS})] = (\bar\beta^2/\beta^2_0)\cdot\Omega(N\log T)$ conditioned on $T = o(\sqrt{N})$. With the threshold $\gamma$ specified as
\begin{equation}
	\gamma_0 = \frac{\log T}{\lambda},
\end{equation}
it follows from Corollary \ref{corollary:1} that
\begin{align}
	\mathbb{P}\big\{|Y|^2\ge\gamma_0\big\}&\ge e^{-\lambda\gamma_0}-\sqrt[4]{2}\rho\xi= \frac{1}{T}-O\bigg(\frac{1}{\sqrt{N}}\bigg)\ge \frac{c}{T}
	\label{CCDF:Y}
\end{align}
with some positive constant $c>0$. The last step in (\ref{CCDF:Y}) is due to the condition that $T=o(\sqrt{N})$. We further extend the above bound to the $T$ random samples:
\begin{align}
    \mathbb{P}\left\{\max_{1\le t\le T}|Y_t|^2\ge \gamma_0\right\}
    &=1-\mathbb{P}\left\{\max_{1\le t\le T}|Y_t|^2< \gamma_0\right\}\notag\\ &=1-\mathbb{P}\left\{|Y_t|^2< \gamma_0\;\text{for all}\; t\right\}\notag\\
    &=1-(1-c/T)^T \ge1-e^{-c}.
\end{align}
According to Markov's inequality, we have
\begin{align}
\mathbb E\left[\max_{1\le t\le T}|Y_t|^2\right]
&\ge \gamma_0\cdot\mathbb{P}\bigg\{\max_{1\le t\le T}|Y_t|^2\ge \gamma_0\bigg\}\notag\\
&\ge \frac{\log T}{\lambda}\cdot (1-e^{-c})\notag\\
&= \Bigg(\sum^N_{n=0}\beta^2_nP+\sigma^2\Bigg)(1-e^{-c})\log T \notag\\
&=\bar\beta^2P\cdot\Omega(N\log T),
	\label{NlogT:lower}
\end{align}
where the expectation is taken over random samples of $\btheta$. The superposition of all the channels from the transmitter to the receiver is denoted as
\begin{equation}
\label{g}
	g = h_0+\sum^N_{n=1}h_ne^{j\theta_n}.
\end{equation}
Let $g_t$ be the value of $g$ for the $t$th random sample and define $t^\star$ as in \eqref{RMS}.
We now explore the relationship between $\mathbb E[\max_{1\le t\le T}|Y_t|^2]$ and $\mathbb E[\max_{1\le t\le T}|g_t|^2]$ as follows:
\begin{align}
	\mathbb E\left[|Y_{t^\star}|^2\right]
	&=\mathbb E\left[|g_{t^\star}X_{t^\star}+Z_{t^\star}|^2\right]\notag\\
	&\le2\mathbb E\left[|g_{t^\star}X_{t^\star}|^2+|Z_{t^\star}|^2\right]\notag\\
	&\le2\mathbb E\bigg[|g_{t^\star}X_{t^\star}|^2+\max_{1\le t\le T}|Z_{t}|^2\bigg]\notag\\
	&=2P\mathbb E\big[|g_{t^\star}|^2\big]+2\mathbb E\bigg[\max_{1\le t\le T}|Z_{t}|^2\bigg]\notag\\
	&=2P\mathbb E\big[|g_{t^\star}|^2\big]+2\sigma^2\Theta(\log T),
	\label{EY:bound}
\end{align}
where the last equality follows by Lemma \ref{lemma:2}. We incorporate the previous bound $\mathbb E[\max_{1\le t\le T}|Y_t|^2]\ge \bar\beta^2P\cdot\Omega(N\log T)$ into (\ref{EY:bound}), obtaining
\begin{align}
	\mathbb E\big[|g_{t^\star}|^2\big]
	&\ge \frac{1}{2}\bar\beta^2\cdot\Omega(N\log T)-\sigma^2\cdot\Theta(\log T)\notag\\
	&=\bar\beta^2\cdot\Omega(N\log T).
\end{align}
It immediately follows that $\mathbb E[f(\btheta^\text{RMS})]=1/\beta^2_0\cdot\mathbb E[|g_{t^\star}|^2]=(\bar\beta^2/\beta^2_0)\cdot\Omega(N\log T)$.

\subsection{Upper Order Bound in (\ref{theorem:random:upper})}
\label{proof:RMS:upper}

We first bound the tail probability of the power of the channel superposition $g$ in (\ref{g}).

\begin{lemma}
\label{lemma:3}
	With each $\theta_n$ drawn from $\Phi_K$ uniformly and independently, the CCDF to the overall channel strength is bounded as
	\begin{equation}
		\mathbb P\big\{|g|^2>\tau\big\}\le 4e^{-\nu\tau/4},
	\end{equation}
	where
	\begin{equation}
	\label{nu}
		\nu = \frac{1}{\sum^N_{n=0}\beta^2_n}
		=\frac{1}{\beta^2_0+N\bar\beta^2}.
	\end{equation}
\end{lemma}
\begin{IEEEproof}
	Let $g_\mathrm{Re}$ be the real part of $g$, so $g_\mathrm{Re}=\mathrm{Re}\{h_0\}+\sum^N_{n=1}\mathrm{Re}\{h_ne^{j\theta_n}\}$. Note that $|\mathrm{Re}\{h_0\}|\le\beta_0$ and $|\mathrm{Re}\{h_ne^{j\theta_n}\}|\le \beta_n$ for each $n$; note also that $\mathrm{Re}\{h_0\},\mathrm{Re}\{h_1e^{j\theta_1}\},\ldots,\mathrm{Re}\{h_Ne^{j\theta_N}\}$ are statistically independent. As a result, $g_\mathrm{Re}$ constitutes a \emph{sub-Gaussian} random variable with the variance proxy $1/\nu$. Then
	\textcolor{black}{
	\begin{equation}
		\mathbb P\big\{g_\mathrm{Re}^2\ge\tau\big\}\le2e^{-\nu\tau/2}.
	\end{equation}}
	By symmetry, the tail probability of the imaginary part of $g$ can be bounded as
	\textcolor{black}{
	\begin{equation}
		\mathbb P\big\{g_\mathrm{Im}^2\ge\tau\big\}\le2e^{-\nu\tau/2}.
	\end{equation}}
	Accordingly, the tail probability of $|g|^2$ can be limited to
	\begin{align}
		\mathbb P\big\{|g|^2\ge\tau\big\}&=\mathbb P\big\{g_\mathrm{Re}^2+g_\mathrm{Im}^2\ge\tau\big\}\notag\\
		&\le \mathbb P\big\{g_\mathrm{Re}^2\ge\tau/2\;\;\text{or}\;\;g_\mathrm{Im}^2\ge\tau/2\big\}\notag\\
		&\le \mathbb P\big\{g_\mathrm{Re}^2\ge\tau/2\big\} + \mathbb P\big\{g_\mathrm{Im}^2\ge\tau/2\big\}\notag\\
		&= 4e^{-\nu\tau/4}.
	\end{align}
	The proof is then completed.
\end{IEEEproof}

We now verify that $\mathbb E\big[f(\btheta^\text{RMS})\big] = (\bar\beta^2/\beta^2_0)\cdot O(N\log T)$ in general. With $ s= \frac{\nu}{8}$ and $t^\star=\arg\max_{1\le t\le T} |Y_t|^2$,
we derive
\textcolor{black}{
\begin{align}
\exp\Big(s\cdot\mathbb E\left[|g_{t^\star}|^2\right]\Big)
&\overset{(a)}\le \mathbb E\left[e^{s|g_{t^\star}|^2}\right]\notag\\
&\le \sum^T_{t=1}\mathbb E\left[e^{s|g_t|^2}\right]\notag\\
& = T\cdot\mathbb E\left[e^{s|g|^2}\right] \notag\\
&\overset{(b)}= T+T\int^\infty_1\mathbb{P}\Big\{e^{s|g|^2}\ge \tau\Big\}d\tau\notag\\
&= T+T\int^\infty_1\mathbb{P}\Big\{|g|^2\ge \frac{\log\tau}{s}\Big\}d\tau\notag\\
&\overset{(c)}\le T+T\int^\infty_1\frac{4}{\tau^2}d\tau\notag\\
&= 5T,
\label{bound:sE}
\end{align}}
where $(a)$ follows by Jensen's inequality, $(b)$ follows by the identity $\mathbb E[X]=d+\int^\infty_{d}\mathbb P\{X\ge x\}dx$ for $X\ge d$, and $(c)$ follows by Lemma \ref{lemma:3}. Furthermore, we take the logarithm of both sides in (\ref{bound:sE}), and thus have \textcolor{black}{$
\mathbb E\left[|g_{t^\star}|^2\right]\le \frac{8}{\nu} \log (5T)= \bar\beta^2\cdot O(N\log T)$}.
With the above result, we can readily bound the SNR boost as $\mathbb E[f(\btheta^\text{RMS})]=(\bar\beta^2/\beta^2_0)\cdot O(N\log T)$. Unlike the lower order bound case, the above proof does not impose any constraint on the number of random samples $T$, so this upper order bound holds in general.

\section{Proof of Theorem \ref{theorem:blind}}
\label{proof:blind}

Recall that $\theta^\text{CSM}_n$ is chosen to maximize $\widehat{\mathbb E}[|Y|^2|\theta_n=k\omega]$ while $\theta^\text{CPP}_n$ is chosen to maximize $\mathbb E[|Y|^2|\theta_n=k\omega]$, so we should have $\theta^\text{CSM}_n=\theta^\text{CPP}_n$ when $\widehat{\mathbb E}[|Y|^2|\theta_n=k\omega]$ and $\mathbb E[|Y|^2|\theta_n=k\omega]$ are sufficiently close to each other for all $k$. We now formalize the above observation. Let $\chi_{n1}$ and $\chi_{n2}$ be the largest and the second-largest values of $\cos(k\omega-\Delta_n)$, respectively, for each $k=1,\ldots,K$; the difference between $\chi_{n1}$ and $\chi_{n2}$ is denoted by $\epsilon_n = \chi_{n1}-\chi_{n2}$.
The key step is to notice that if
\begin{equation}
\label{AppendixB:epsilon}
	\Big|\widehat{\mathbb E}[|Y|^2|\theta_n=k\omega]-\mathbb E[|Y|^2|\theta_n=k\omega]\Big|<2\beta_0\beta_nP\epsilon_n
\end{equation}
for every $k$, then we must have
$
\arg\max_{\varphi\in\Phi_K}\widehat{\mathbb E}[|Y|^2|\theta_n=\varphi]
	=\arg\max_{\varphi\in\Phi_K}{\mathbb E}[|Y|^2|\theta_n=\varphi]$,
and therefore $\theta^\text{CSM}_n=\theta^\text{CPP}_n$. In the remainder of the proof, we discuss in what regime of $(N,T)$ the condition (\ref{AppendixB:epsilon}) holds for all $n=1,\ldots,N$.


Without loss of generality, we focus on a particular $(n,k)$ and its corresponding conditional subset $\mathcal Q_{nk}$. Let $T_{nk}$ be the cardinality of $\mathcal Q_{nk}$. With each $\theta_{nt}$ drawn from $\Phi_K$ uniformly and independently, we have 
\begin{equation}
\label{LLN}
T_{nk}=\frac{T}{K}\quad\text{with high probability}.
\end{equation}
Consider the noise free received signal $Y_t-Z_t$. We denote its expected power conditioned on $\theta_n=k\omega$ as
\begin{subequations}
\begin{align}
	\eta_{nk} 
	&= \mathbb E\left[|Y_t-Z_t|^2|\theta_n=k\omega\right]\\
	&= P\big|h_0+h_ne^{jk\omega}\big|^2+\sum\nolimits_{m\ne n}\beta_m^2P,
\end{align}
\end{subequations}
while denoting the corresponding conditional sample mean as
\begin{equation}
	\hat\eta_{nk} = \frac{1}{T_{nk}}\sum_{t\in\mathcal Q_{nk}}|Y_t-Z_t|^2.
\end{equation}
We denote the conditional sample mean of noise power as
\begin{equation}
	\hat\sigma^2_{nk} = \frac{1}{T_{nk}}\sum_{t\in\mathcal Q_{nk}}|Z_t|^2.
\end{equation}
Moreover, we define the conditional sample mean of a cross term to be
\begin{equation}
\delta_{nk}=\frac{2}{T_{nk}}\sum_{t\in\mathcal Q_{nk}}\mathrm{Re}\left\{(Y_t-Z_t)^*Z_t\right\}.
\end{equation}
Most importantly, notice that
\begin{subequations}
\label{sum_statistics}
\begin{align}
\mathbb E[|Y|^2|\theta_n=k\omega]
&= \eta_{nk} + \sigma^2_{nk},\\
\widehat{\mathbb E}[|Y|^2|\theta_n=k\omega]
&= \hat\eta_{nk} + \hat\sigma^2_{nk} + \delta_{nk}.
\end{align}
\end{subequations}
We consider the four error events for some $q>0$ and $\epsilon>0$:
\begin{subequations}
\begin{align}
	\mathcal E_{nk,1}(q)&=\big\{\eta_{nk}\ge qP/\nu\big\},\\
	\mathcal E_{nk,2}(\epsilon)&=\big\{\big|\hat\eta_{nk}-\eta_{nk}\big|\ge \epsilon\big\},\\
	\mathcal E_{nk,3}(\epsilon)&=\big\{\big|\hat\sigma^2_{nk}-\sigma^2\big|\ge\epsilon\big\},\\
	\mathcal E_{nk,4}(\epsilon)&=\big\{|\delta_{nk}|\ge\epsilon\big\},
\end{align}
\end{subequations}
where $\nu$ is defined in (\ref{nu}). The probability of each of the above error events is upper bounded in what follows. According to Lemma \ref{lemma:3}, we immediately have
\begin{equation}
\label{bound:E1}
	\mathbb P\{\mathcal E_{nk,1}(q)\}\le 4e^{-q/4}.
\end{equation}
Conditioned on $\mathcal E^c_{nk,1}$, i.e., assuming that
\begin{align}
	0\le|Y_t-Z_t|^2< qP/\nu\;\;\text{for all}\;\; t\in\mathcal Q_{nk},
\end{align}
we obtain the following bound from Hoeffding's inequality:
\begin{align}
	\mathbb P\left\{\mathcal{E}_{nk,2}(\epsilon)|\mathcal{E}^c_{nk,1}(q)\right\}\le 2\exp\left(-\frac{2\epsilon^2\nu^2T_{nk}}{q^2P^2}\right).
\end{align}
The following bound follows by Chebyshev's inequality:
\begin{align}
	\mathbb P\big\{\mathcal E_{nk,3}(\epsilon)\big\}
	\le \frac{7\sigma^4}{\epsilon^2T_{nk}}.
\end{align}
By virtue of Chebyshev's inequality, we also have
\begin{align}
\label{bound:E4}
	\mathbb P\big\{\mathcal E_{nk,4}(\epsilon)|\mathcal E^c_{nk,1}(q)\big\}
	\le \frac{2qP\sigma^2}{\nu\epsilon^2T_{nk}}.
\end{align}
The purpose of developing the above inequalities \eqref{bound:E1}--\eqref{bound:E4} is to bound the probability of the following error event from above:
\begin{equation}
\mathcal{E}_{nk,0} = \Big\{\Big|\widehat{\mathbb E}[|Y|^2|\theta_n=k\omega]-{\mathbb E}[|Y|^2|\theta_n=k\omega]\Big|\ge \epsilon_0\Big\},
\end{equation}
where the constant $\epsilon_0 = \min_{n=1,\ldots,N}\left\{2\beta_0\beta_n\epsilon_n\right\}$.
We show that
\allowdisplaybreaks
\begin{align}
	&\mathbb P\{\mathcal{E}_{nk,0}\}\notag\\
	&\overset{(a)}=\mathbb P\big\{\big|\hat\eta_{nk}+\hat\sigma^2_{nk}+\delta_{nk}-\eta_{nk}-\sigma^2\big|\ge\epsilon_0\big\}\notag\\
	&\overset{(b)}\le\mathbb P\left\{\big|\hat\eta_{nk}-\eta_{nk}\big|\ge\frac{\epsilon_0}{3}\right\}+\mathbb P\left\{\big|\hat\sigma^2_{nk}-\sigma^2\big|\ge\frac{\epsilon_0}{3}\right\}\notag\\
	&\qquad+\mathbb P\left\{\big|\delta_{nk}\big|\ge\frac{\epsilon_0}{3}\right\}
	\notag\\
	&=\mathbb P\big\{\mathcal E_{nk,2}(\epsilon_0/3)\big\}+\mathbb P\big\{\mathcal E_{nk,3}(\epsilon_0/3)\big\}+\mathbb P\big\{\mathcal E_{nk,4}(\epsilon_0/3)\big\}\notag\\
	&\le\mathbb P\big\{\mathcal E_{nk,2}(\epsilon_0/3)|\mathcal E^c_{nk,1}(q)\big\}+\mathbb P\big\{\mathcal E_{nk,3}(\epsilon_0/3)|E^c_{nk,1}(q)\big\}\notag\\
	&\qquad+\mathbb P\big\{\mathcal E_{nk,4}(\epsilon_0/3)\big\}+ 2\mathbb P\big\{\mathcal E_{nk,1}(q)\big\}\notag\\
	&\overset{(c)}\le2\exp\left(-\frac{2\epsilon^2_0\nu^2T}{9q^2P^2K}\right)+\frac{63\sigma^4K}{\epsilon^2_0T}
	+\frac{18qP\sigma^2K}{\nu\epsilon^2_0T}+8e^{-q/4}\notag\\
	&\qquad \text{with high probability,}
\end{align}
where $(a)$ follows by \eqref{sum_statistics}, $(b)$ follows since $|a_1+a_2+a_3|\ge\epsilon_0$ implies that at least one $|a_i|\ge\epsilon_0/3$, and $(c)$ follows by applying \eqref{LLN} and \eqref{bound:E1}--\eqref{bound:E4} jointly.

Furthermore, we consider the overall error event
$\mathcal E_0 = \bigcup_{(n,k)}\mathcal E_{nk,0}$,
the probability of which can be upper bounded by the union of the events bound as
\begin{align}
&\mathbb P\{\mathcal E_0\}\notag\\
&\le \sum^K_{k=1}\sum^{N}_{n=1}\mathbb P\{\mathcal E_{nk,0}\}\notag\\
&\le 2NK\exp\left(-\frac{2\epsilon^2_0\nu^2T}{9q^2P^2K}\right)+\frac{63\sigma^4NK^2}{\epsilon^2_0T}+\frac{18qP\sigma^2NK^2}{\nu\epsilon^2_0T}
\notag\\
&\qquad+8NKe^{-q/4}\;\; \text{with high probability}.
\end{align}
For\footnote{Actually, we just need to bound $p_0$ away from $\frac14$.} any $0< p_0\le \frac15$, we have
\begin{align*}
	2NK\exp\left(-\frac{2\epsilon^2_0\nu^2T}{9q^2P^2K}\right)&\le p_0\;\;\text{if}\;\;T=\Omega(N^2q^2\log N),\\
	\frac{63\sigma^4NK^2}{\epsilon^2_0T}&\le p_0\;\;\text{if}\;\;T=\Omega(N),\\
	\frac{18qP\sigma^2NK^2}{\nu\epsilon^2_0T}&\le p_0\;\;\text{if}\;\;T=\Omega(N^2q),\\
	8NKe^{-q/4}&\le p_0\;\;\text{if}\;\;q=\Omega(\log N).
\end{align*}
Combining the above results, we obtain that $\mathbb P\{\mathcal E_0\}\le 4p_0$ and hence $\mathbb P\{\mathcal E^c_0\}\ge1-4p_0\ge\frac{1}{5}$ whenever $T=\Omega(N^2(\log N)^3)$. Consequently,
$\mathbb E[f(\btheta^\text{CSM})]\ge f(\btheta^\text{CPP}) \cdot\mathbb P\{\mathcal E^c_0\}=(\bar\beta^2/\beta^2_0)\cdot \Omega(N^2)$.
The above lower upper bound together with the upper order bound in Proposition \ref{prop:upper_bound} gives $\mathbb E[f(\btheta^\text{CSM})]=(\bar\beta^2/\beta^2_0)\cdot \Theta(N^2)$.

\section{Proof of Theorem \ref{theorem:enhanced_blind}}
\label{proof:ECSM}

In Appendix \ref{proof:blind} we prove that CSM achieves the optimal SNR boost when $T=\Omega(N^2(\log N)^3)$ for $K\ge3$. Since ECSM is a strengthened version of CSM, we readily have 
$\mathbb E[f(\btheta^\text{ECSM})]=(\bar\beta^2/\beta^2_0)\cdot \Theta(N^2)$ when $T=\Omega(N^2(\log N)^3)$ for $K\ge3$. The rest of the proof concentrates on the case of $K=2$.

As shown in Remark \ref{ECSM:K2}, when $K=2$, ECSM decides each phase shift $\theta_n$ according to the sign of $\mathrm{Im}\big\{\widehat{\mathbb E}[Y|\theta_n=\theta_n'']\big\}$. In principle, if $\mathrm{Im}\{\widehat{\mathbb E}[Y|\theta_n=\theta''_n]\}$ and $\mathrm{Im}\{\mathbb E[Y|\theta_n=\theta''_n]\}$ are of the same sign for all possible $\theta''_n$, we can recover the solution in \cite{IRS:opt_21} as 
$\btheta^\text{ECSM}$. Thus, we are interested in the error event
\begin{multline}
	\mathcal E_{nk}=\big\{\mathbbm{1}_{+}\big(\mathrm{Im}\{\widehat{\mathbb E}[Y|\theta_n=k\omega]\}\big)\ne\\
	\mathbbm{1}_{+}\big(\mathrm{Im}\{\mathbb E[Y|\theta_n=k\omega]\}\big)\big\}.
\end{multline}
Without loss of generality, let $X=\sqrt{P}$. Define
\begin{align}
	\psi_{nk}&=\mathrm{Im}\{{\mathbb E}[Y|\theta_n=k\omega]\}\notag\\
	&=\sqrt{P}\beta_0\sin(\alpha_0) + \sqrt{P}\beta_n\sin(\alpha_n+k\omega).
\end{align}
Let $T_{nk}$ be the cardinality of the conditional subset $\mathcal Q_{nk}$. We can then bound the error probability as
\begin{align}
\mathbb P\{\mathcal E_{nk}\}
&\le\mathbb P\Big\{\Big|\mathrm{Im}\{\widehat{\mathbb E}[Y|\theta_n=k\omega]\}-\psi_{nk}\Big|\ge|\psi_{nk}|\Big\}\notag\\
&= \mathbb P\left\{\left|\frac{1}{T_{nk}}\sum_{t\in\mathcal Q_{nk}}\mathrm{Im}\big\{Y_t\big\}-\psi_{nk}\right|\ge|\psi_{nk}|\right\}\notag\\
&\le \mathbb P\left\{\left|\frac{1}{T_{nk}}\sum_{t\in\mathcal Q_{nk}}\mathrm{Im}\{Y_{t}-Z_t\}-\psi_{nk}\right|\ge \frac{|\psi_{nk}|}{2}\right\}\notag\\
&\qquad + \mathbb P\left\{\left|\frac{1}{T_{nk}}\sum_{t\in\mathcal Q_{nk}}\mathrm{Im}\{Z_t\}\right|\ge\frac{|\psi_{nk}|}{2}\right\}.
\end{align}
Because $\mathrm{Im}\{Z_t\}\;\text{i.i.d.}\;\sim\mathcal{N}(0,\sigma^2/2)$, we obtain the following bound from Chebyshev's inequality:
\begin{equation}
	\mathbb P\left\{\left|\frac{1}{T_{nk}}\sum_{t\in\mathcal Q_{nk}}\mathrm{Im}\{Z_t\}\right|>\frac{|\psi_{nk}|}{2}\right\}\le \frac{2\sigma^2}{T_{nk}\psi^2_{nk}}.
\end{equation}
Notice that $\big|\mathrm{Im}\{Y_t-Z_t\}\big|\le \sqrt{P}\sum^N_{m=0}\beta_m$. In light of Hoeffding's inequality, we have
\begin{align}
&\mathbb P\left\{\left|\frac{1}{T_{nk}}\sum_{t\in\mathcal Q_{nk}}\mathrm{Im}\{Y_t-Z_t\}-\psi_{nk}\right|\ge \frac{|\psi_{nk}|}{2}\right\}\notag\\
&\le 2\exp\left(-\frac{2(\psi_{nk}/2)^2T_{nk}}{4P(\sum^N_{m=0}\beta_m)^2}\right)\notag\\
&\le 2\exp\left(-\frac{\psi_{nk}^2T_{nk}\nu}{16PN}\right)\notag\\
&= 2\exp\left(-\frac{\psi_{nk}^2T\nu}{16PNK}\right)\;\text{with high probability},
\end{align}
where the second inequality follows by $\big(\sum_ix_i\big)^2\le2\sum_{i}x^2_i$.
The overall error event is defined as
$\mathcal E = \bigcup_{(n,k)}\mathcal E_{nk}$.
The overall error probability can be bounded as
\begin{align}
	\mathbb P\{\mathcal E\}
	&\le \sum^K_{k=1}\sum^N_{n=1}\mathbb P\{\mathcal E_{nk}\}\notag\\
	&\le 2NK\exp\left(-\frac{\psi_0^2T\nu}{16PNK}\right) + \frac{2NK^2\sigma^2}{T\psi_{0}},
\end{align}
where $\psi_{0}=\min_{n,k}\psi_{nk}$. For any $0<p_0\le\frac{1}{4}$, we have
\begin{align*}
	2NK\exp\left(-\frac{\psi_0^2T\nu}{16PNK}\right)&\le p_0\;\;\text{if}\;\;T=\Omega(N^2\log N),\\
	\frac{2NK^2\sigma^2}{T\psi_{0}}&\le p_0\;\;\text{if}\;\;T=\Omega(N).
\end{align*}
Thus, we can ensure that $\btheta^\text{ECSM}$ recovers the enhanced solution in \cite{IRS:opt_21} with probability $\mathbb P\{\mathcal E^c\}\ge 1-2p_0\ge \frac{1}{2}$ when $T=\Omega(N^2\log N)$. As a result, $\mathbb E[f(\btheta^\text{ECSM})]$ has the same order bound as the enhanced algorithm \cite[Theorem~2]{IRS:opt_21} for any $K\ge2$, i.e., $\mathbb E[f(\btheta^\text{ECSM})]=\frac{\bar\beta^2}{\beta^2_0}\cdot \Theta(N^2)$. 

\bibliographystyle{IEEEbib}
\bibliography{IEEEabrv,strings}

\begin{IEEEbiographynophoto}{Shuyi Ren}
	(Student Member, IEEE)
	received the B.S. degree in applied mathematics from Tongji University, Shanghai, China, in 2018. She is currently working toward the Ph.D. degree in computer \& information engineering in the School of Science and Engineering at The Chinese University of Hong Kong (Shenzhen), Shenzhen, China. Her research interests include signal processing, statistics, machine learning and optimization algorithms.
\end{IEEEbiographynophoto}

\begin{IEEEbiographynophoto}{Kaiming Shen}
(Member, IEEE) received the B.Eng. degree in information security and the B.S. degree in mathematics from Shanghai Jiao Tong University, Shanghai, China in 2011, then the M.A.Sc. and Ph.D. degrees in electrical and computer engineering from University of Toronto, Ontario, Canada in 2013 and 2020, respectively.

Since 2020, he has been an Assistant Professor with the School of Science and Engineering at The Chinese University of Hong Kong (Shenzhen), China. His main research interests include optimization, wireless communications, and information theory. Dr. Shen received the IEEE Signal Processing Society Young Author Best Paper Award in 2021.
\end{IEEEbiographynophoto}

\begin{IEEEbiographynophoto}{Yaowen Zhang}
	(Student Member, IEEE)
	received the B.S. degree in applied physics from Beijing Institute of Technology, Beijing, China, in 2020. He is currently working toward the M.Phil. degree in the computer \& information engineering in the School of Science and Engineering at The Chinese University of Hong Kong (Shenzhen), Shenzhen, China. His research interests include signal processing and optimization algorithms.
	\end{IEEEbiographynophoto}

\begin{IEEEbiographynophoto}{Xin Li}
graduated from Xidian University and joined Huawei in 2008. He has rich experience in wireless channel modeling and wireless network performance modeling and optimization. Currently, he is a technical expert in Huawei's experience lab, focusing on future-oriented network technology research, including new technologies such as Intelligent Reflection Surface and Intelligent Transmission Surface, and their application in network structure optimization.
\end{IEEEbiographynophoto}

\begin{IEEEbiographynophoto}{Xin Chen}
graduated from the Radio Engineering Department of Southeast University and joined Huawei in 2000. He has 20 years of R\&D experience in the wireless communications field. He has served as senior algorithm engineer, solution architect, and technology development director successively. He has rich experience and achievements in network planning, optimization, and operation and maintenance of mobile communication networks.

Currently, he is the director of the Algorithm Dept of the Service and Software R\&D domain of Huawei Carrier BG. He is responsible for the research of key technologies for digitalization and intelligence in the telecom field, including intelligent network optimization, autonomous network architecture and O\&M, analysis algorithms of telecom big data, and next-generation computing architecture in the telecom field.
\end{IEEEbiographynophoto}

\begin{IEEEbiographynophoto}{Zhi-Quan Luo}
	(Fellow, IEEE)
 received the B.S. degree in applied mathematics from Peking University, China, and the Ph.D. degree in operations research from MIT in 1989. From 1989 to 2003, he held a faculty position with the ECE Department of McMaster University, Canada. He held a tier-1 Canada Research Chair in information processing from 2001 to 2003. After that, he has been a full professor at the ECE Department, University of Minnesota and held an endowed ADC Chair in digital technology. Currently, he is the Vice President (Academic) of The Chinese University of Hong Kong (Shenzhen) and the director of Shenzhen Research Institute of Big Data (SRIBD). Prof. Luo is a Fellow of IEEE and SIAM. He was elected to Fellow of Royal Society of Canada in 2014 and a Foreign Member of the Chinese Academy of Engineering (CAE) in 2021. He received four best paper awards from the IEEE Signal Processing Society, one best paper award from EUSIPCO, the Farkas Prize from INFORMS and the prize of Paul Y. Tseng Memorial Lectureship in Continuous Optimization as well as some best paper awards from international conferences. In 2021, he was awarded 2020 ICCM Best Paper Award by International Consortium of Chinese Mathematicians. He has published over 350 refereed papers, books and special issues. He was the Editor-in-Chief for IEEE Transactions on Signal Processing(2012-2014) and served as the Associate Editor for many internationally recognized journals. His research mainly addresses mathematical issues in information sciences, with particular focus on the design, analysis and applications of large-scale optimization algorithms. 
\end{IEEEbiographynophoto}







\end{document}